\PassOptionsToPackage{table}{xcolor}

\documentclass[unnumsec,webpdf,modern,large]{oup-authoring-template}%

\usepackage{natbib}
\usepackage{tabularx}

\theoremstyle{thmstyleone}%
\theoremstyle{thmstyletwo}%
\theoremstyle{thmstylethree}%

\usepackage{csquotes}
\usepackage{booktabs}
\usepackage{comment}
\usepackage{threeparttable}
\usepackage{longtable}
\usepackage{hyperref}

\journaltitle{Submitted to [Omitted] journal}
\copyrightyear{2025}
\pubyear{2025}
\appnotes{Preprint Paper}

\let\clearemptydoublepage\clearpage

\makeatletter
\def\@oddfoot{}
\def\@evenfoot{}
\makeatother

\begin{document}

\title{Can Large Language Models Automate Phishing Warning Explanations? A Controlled Experiment on Effectiveness and User Perception}

\author[1]{Federico Maria Cau}
\author[2]{Giuseppe Desolda} 
\author[2]{Francesco Greco}
\author[1]{Lucio Davide Spano}
\author[3]{Luca Viganò}

\address[1]{University of Cagliari, Cagliari 09124, Italy}
\address[2]{Department of Computer Science, University of Bari Aldo Moro, Bari 70125, Italy}
\address[3]{Department of Informatics, King's College London, London WC2B 4BG, UK}

\corresp{\textsuperscript{*}Corresponding author. E-mail: giuseppe.desolda@uniba.it}

\abstract{
Phishing has become a prominent risk in modern cybersecurity, often used to bypass technological defences by exploiting predictable human behaviour. Warning dialogues are a standard mitigation measure, but the lack of explanatory clarity and static content limits their effectiveness. In this paper, we report on our research to assess the capacity of Large Language Models (LLMs) to generate clear, concise, and scalable explanations for phishing warnings. 
We carried out a large-scale between-subjects user study (N = 750) to compare the influence of warning dialogues supplemented with manually generated explanations against those generated by two LLMs, Claude 3.5 Sonnet and Llama 3.3 70B. We investigated two explanatory styles (feature-based and counterfactual) for their effects on behavioural metrics (click-through rate) and perceptual outcomes (e.g., trust, risk, clarity). 
The results provide empirical evidence that LLM-generated explanations achieve a level of protection statistically comparable to expert-crafted messages, effectively automating a high-cost task. While Claude 3.5 Sonnet showed a trend towards reducing click-through rates compared to manual baselines, Llama 3.3, despite being perceived as clearer, did not yield the same behavioral benefits. Feature-based explanations were more effective for genuine phishing attempts, whereas counterfactual explanations diminished false-positive rates. 
Other variables, such as workload, gender, and prior familiarity with warning dialogues, significantly moderated the effectiveness of warnings. These results indicate that LLMs can be used to automatically build explanations for warning users against phishing, and that such solutions are scalable, adaptive, and consistent with human-centred values.
}


\maketitle 

\section{Introduction}
\label{sec1}

Phishing attacks are among the most widespread and harmful cyber threats, leveraging human behaviour instead of technical weaknesses~\citep{Khonji2013Phishing, Prakash2010PhishNet}. There is a vast number of automatic filtering techniques (e.g., blacklists, machine learning-based detection), but these automatic defence systems are far from perfect~\citep{Basit2021Survey}, and there is a strong possibility of false positives and, even more importantly, false negatives, which could result in users receiving harmful material.

To overcome this issue, most defensive systems engage the human in the decision-making process by presenting users with \textit{warning dialogues}, allowing them to make the final decision and possibly prevent a phishing attack~\citep{desolda2019alerting}. Although warning dialogues are commonly used, they are effective only when clearly describing the risk and influencing the decision to act upon it~\citep {ibm2024security}. Most dialogues used today are passive and mostly static, with no clear explanations as to why a particular message is harmful~\citep{egelman2008warned, wu2006security}. 

In the last few years, researchers working in \textit{Human-Centred Security} and \textit{Explainable AI (XAI)} have investigated whether feature-based explanations might be effective in mitigating phishing attacks~\citep{Guidotti2019Survey, Gunning2019DARPAs}. 
The clear advantage of such explanations is the reduction of susceptibility to phishing, but they are hand-made, expensive to maintain, and might not easily adjust to the changing circumstances of the threats~\citep{desolda2023explanations, buono2023warnings, greco2025enhancing}. 
Other ways of generating explanations have also been considered in general, such as the use of counterfactuals~\citep{Guidotti2019Counterfactual} or the use of \textit{Large Language Models (LLMs)}, but they have not been investigated extensively in the specific field of cybersecurity~\citep{Xu2025LLMsCybersecurity}.

To bridge these gaps, we have carried out a large-scale, between-subjects user study ($N = 750$). 
We investigated how well phishing warning dialogues with explanations can be generated with LLMs. We assessed the result of using a commercial closed-weight model (Claude 3.5 Sonnet) and an open-weight alternative (Llama 3.3 70B). Additionally, we considered two styles of explanation: feature-based and counterfactual explanations. A baseline with feature-based, manually written explanations has also been considered ~\citep{desolda2023explanations, greco2025enhancing}. The effectiveness of the different experimental conditions has been evaluated by considering both measures of behaviour (e.g., click-through rate on phishing links) and user perceptions (e.g., clarity, trust, perceived risk), and by modelling the degree to which individual factors, including cognitive style and knowledge of cybersecurity issues, influence results.

We found that explanations generated by LLMs can be as effective as those produced manually; specifically, explanations generated by Claude were better than those crafted manually. Regarding the type of explanation, counterfactual ones have an interesting power in preventing false positive cases, while feature-based ones have certain benefits in identifying true positive cases. Moreover, user-related factors such as gender, familiarity, and mental workload significantly influence the effectiveness of the warning.

The remainder of this paper is structured as follows.  
Section~\ref{sec:related_work} reviews related work on phishing defences, warning dialogue design, and explainable AI in cybersecurity.  
Section~\ref{sec:approach} presents the experimental design and the explanation generation process. This section also reports the formalisation of the four research questions this study aims to answer. 
Section~\ref{sec:methods} details the study procedure and methodology.  
Section~\ref{sec:results} reports the quantitative and qualitative results addressing our four research questions. Section~\ref{sec:threats_validity} discusses threats to validity.  
Section~\ref{sec:lessons_learned} synthesises the findings as a set of lessons learned. Section ~\ref{sec:conclusions} concludes the paper by summarising the main results and discussing potential areas for future research.

\section{Related work} \label{sec:related_work}

\subsection{Warning dialogues for Phishing Defence}

The complete elimination of phishing through automated filtering methods such as blocklists~\citep{gupta2014emerging} and AI-based methods \cite{Basit2021Survey} is currently unfeasible. No existing tools can detect phishing content, such as websites or emails, with perfect accuracy~\citep{elassal2020benchmarking, gholampour2023adversarial}. Furthermore, attempts at reducing the number of false negatives (i.e., phishing emails that go undetected) tend to increase the number of false positives (i.e., legitimate emails misclassified as phishing) in classification tasks \citep{saxena2018precision}. This eventually leads to filtering out genuine, relevant emails, potentially disrupting user productivity. 

The solution currently adopted in most systems is not to automatically filter suspicious content classified as phishing with low confidence; instead, these emails are presented to users alongside a warning dialogue that alerts them to possible threats, leaving the final decision to the users \citep{kumaraguru2010teaching}. Despite warning dialogues, many users fail to grasp the significance or meaning of these messages, so phishing attacks remain highly effective \citep{ibm2024security}.

One key problem of everyday warning dialogues is their passive behaviour. Popular email clients such as Gmail implement static toolbars above the email content. Warnings like these are largely ineffective, as users often overlook or ignore them \citep{egelman2008warned, wu2006security}. 
On the other hand, warnings that implement an active behaviour, i.e., interrupt user interactions and demand attention, have been proven to be significantly more effective \citep{wogalter2002research, petelka2019put, buono2023warnings, egelman2008warned, greco2025enhancing}. However, such warnings are currently implemented only in web browsers \citep{desolda2019alerting}, despite the benefits of active behaviour having also been demonstrated for warnings shown in email clients \citep{greco2025enhancing, buono2023warnings}.

Another limitation of warnings is the phenomenon known as \textit{habituation}. This behaviour occurs when users are repeatedly exposed to the same warning stimulus, even under varying risk conditions, leading to decreased attentional responses and increased likelihood of bypassing the warning \citep{akhawe2013alice, kim2009habituation}. To mitigate this problem, \textit{polymorphic} warnings were demonstrated to be significantly more resistant to habituation than static warnings \citep{anderson2015polymorphic}. Such warnings are called polymorphic since they dynamically change their appearance or content to provide varied visual stimuli. Nonetheless, they are still not implemented in browsers or email clients despite their advantages \citep{desolda2019alerting}. 
Only a few studies have explored the use of polymorphic warnings paired with tailored explanations to enhance their effectiveness \citep{desolda2023explanations, buono2023warnings}, or visual nudges based on social salience to enhance effectiveness \citep{nicholson2017can}. \citeauthor{desolda2023explanations} found that including manually designed explanations of malicious features in warning dialogues in web browsers led to warnings that were more familiar and understandable than those in the most common browsers. \citeauthor{buono2023warnings} studied warnings with explanations in the context of email clients and found that active warnings with explanations based on phishing features largely outperformed contextual warnings without explanations. This study confirmed the importance of interrupting user interaction flow to focus entirely on the message, similar to active warnings shown in browsers. Additionally, the study demonstrated the importance of explaining to users why content is malicious, helping them make better-informed decisions.

A further issue that compromises the effectiveness of the warnings is the absence of detailed explanations about the phishing risk \citep{bravolillo2011improving}. Ideally, warnings should communicate clearly why specific content is dangerous and outline the potential consequences of ignoring it. However, many current warnings provide vague messages that fail to identify the malicious elements within emails or websites. This places the burden of risk assessment on the users, who often lack cybersecurity expertise or additional information to make informed decisions \citep{desolda2021human}. This is particularly critical in lateral phishing scenarios, where users struggle to detect subtle changes in message content and over-rely on sender identity, which appears legitimate \cite{chitare2023may}.
By incorporating explanations into warnings, the users’ understanding of the threat can be improved, addressing knowledge gaps, and potentially increasing compliance and trust in the system \citep{bravolillo2011improving, RN3}. This aligns with findings suggesting that transparent communication about incidents is crucial for building trust and understanding between security practitioners and employees \citep{chitare2025exploring}.

The use of explanatory warnings in web browsers has been explored in \citep{desolda2023explanations}. The authors identified seven website features that can be intuitively explained to users and produced two explanation messages for each. These messages were refined through an iterative process to optimise text readability and sentiment. Moreover, the explanation messages followed a well-defined structure grounded in warning theory \citep{bauer2013warning}, which led to the identification of a template for generating explanation messages: ``Feature description + Hazard Explanation + Consequences of not complying with the warning''. 
The authors empirically validated the adoption of warnings with explanations among 150 users, comparing it with warnings without explanations. Results showed that the proposed warnings were clearer and more effective than those implemented in modern browsers, which lack explanations. \cite{greco2025enhancing} explored a similar approach in the context of email clients, combining active warning dialogues with embedded explanations. The findings indicated that this combination of active behaviour and explanation significantly outperformed the state-of-the-art email client warnings proposed by \cite{petelka2019put}.

As security software engineers, we must understand the underlying decision-making algorithm to generate explanations. In the case of AI-based phishing filtering methods, explanations can be produced by employing \emph{eXplainable AI (XAI)} techniques \citep{Guidotti2019Survey,Gunning2019DARPAs}.\footnote{See \citep{ViganoMagazzeni2020} for a discussion on the relationship between explainable cybersecurity and explainable AI.} These allow one to explain the behaviour and decisions of AI models, which are often black-box, and thus not directly interpretable even by AI engineers. 
With XAI techniques, for example, the system can provide a rationale for the reasons that led to the automatic classification of an email as phishing. 
Explanations in warning dialogues can contain such a rationale to improve the transparency of the AI system. Let us explore these aspects in the following subsection. 

\subsection{Human-Centred Explainable AI for Phishing Attacks}
\label{sec:hcxai_phishing}
Integrating XAI solutions into cybersecurity systems is crucial for maintaining user trust and confidence. This approach contrasts with other human-centric interventions, such as simulated phishing campaigns, which have been argued to potentially damage the trust relationship between employees and security teams \cite{Murdoch2017}. This underscores the importance of user-centred experimental evaluations that balance security, usability, and resilience to adversarial attacks \citep{Charmet2022XAI4Cybersecurity, Rjoub2023XAI4Cybersecurity, SARKER2024XAI4CS, Samed2025XAI4CSIntrusionDetectionSystems}.
In this regard, the broader \emph{Human-Centred XAI (HCXAI)} literature offers valuable insights into multiple explanation types and their effects on users.
The most prevalent explanations in AI-assisted decision systems are feature-based ones \citep{Lai2023SurveyHumanAIDecisionMaking}, which clarify \textit{why} an AI system produced a specific decision by highlighting the most important features driving the decision.
Feature-based explanations may enhance user understanding, awareness of uncertainty, trust calibration, and decision accuracy, although they can sometimes foster overreliance on AI \citep{Zhang2020ConfidenceExplanationsAccuracyTrust,Wang2021ExampleBasedFeatureBasedAndOthers,Chen2023RelianceExampleBasedFeatureBased,Ma2023CorrectnessLikelihoodAIUsers,Fok24FeatureBased,Cau25MultiExp}. 
Counterfactual explanations constitute another, less explored style: they provide not only \textit{why} an AI system reached a specific decision, but also \textit{how} it is possible for users to alter input features to achieve a potential desired outcome by presenting contrastive \textit{``what-if''} statements. While they are mainly used to offer actionable recourse when negative decisions occur---e.g., loan denial---by indicating necessary feature changes for a positive result---e.g., increasing income for approval---\citep{Koh24AlgorithmicRecourse,Verma24AlgorithmicRecourse,VanNostrand24CounterfactualActionableRecourse,Upadhyay25AlgorithmicRecourse,Cau2025HCXAILoan}, they have also demonstrated improvements in helpfulness, decision alignment, and accuracy, sometimes matching or exceeding feature-based methods, in AI-assisted decision-making tasks \citep{Scharowski2023FeatureImportanceCounterfactualsExplanations, Celar2023CounterfactualCausal, Cau2023LogicalReasoningStock, Teso2023InteractiveExplanations, Lee23Counterfactual, Gentile25Counterfactual}.

Regarding XAI in cybersecurity and phishing in particular, only a small number of proof-of-concept studies have assessed the effectiveness of explanations with real users \citep{desolda2023explanations,greco2025enhancing}.
For example, \cite{desolda2023explanations} found that adding a feature-based explanation to a phishing warning dialogue made the warning more understandable and familiar, which may deter users from visiting malicious sites. In a similar study, \citet{greco2025enhancing} compared warnings with and without feature-based explanations, as well as the effect of displaying the warning before versus after opening an email. They demonstrated that adding explanations enhanced users' comprehension of the phishing threat and, when the warning appeared after an email was opened, resulted in a measurable reduction in click-through rate (CTR) despite the additional effort required by participants.
However, these investigations have focused exclusively on manually crafted feature‐based explanations, leaving the generation of feature‐based or alternative explanation types via large language models largely unexplored \citep{SARKER2024XAI4CS}.

This work addresses these gaps through a user study in a phishing email scenario. We compare LLM-generated, feature-based, and counterfactual explanations to manually crafted ones within warning dialogues, and evaluate their impact on users' CTR on potentially harmful links and subjective perceptions of the warning dialogue.

\section{Approach} \label{sec:approach}

In this study, we focus on the impact of two factors on user decision-making in phishing-warning dialogues: (i) the style of the explanation messages in the warning dialogue and (ii) the LLM employed to generate the explanation message. 

Regarding the explanation style, the first approach we considered is based on \textit{feature-based explanations}, a widely adopted approach in HCI systems for communicating the most influential features contributing to an AI decision \citep{Lai2023SurveyHumanAIDecisionMaking,Fok24FeatureBased}. In our context, these explanations highlight specific elements of an email, such as a masked link or a suspicious sender address, that led the AI to classify it as phishing. Complementing this standard, we also include \textit{counterfactual explanations}, which offer an alternative, theoretically grounded approach to supporting user understanding. Rather than simply highlighting influential features, counterfactuals offer contrastive explanations answering implicit \textit{\enquote{why not?}} questions by describing how small, meaningful changes would alter the AI’s prediction \citep{Lee23Counterfactual,Koh24AlgorithmicRecourse,Gentile25Counterfactual}. This explanation technique clarifies decision boundaries and supports actionable recourse, particularly relevant in high-stakes settings such as fraud detection or security. In our phishing warning dialogues, we embed these explanations to show how the email could have avoided being flagged, for example, stating that the email would be considered safe if the link label matched the actual destination URL (e.g., \enquote{protect your account} linking to facebook.com/protect-account).

The second factor concerns the LLM model used to generate the explanation message. While past studies often relied on manually written explanation messages to demonstrate their effectiveness \citep{desolda2023explanations, greco2025enhancing, buono2023warnings}, the growing availability of LLMs may enable the automatic generation of tailored content in both explanation styles. 
If explanations produced by LLMs prove as effective as manually crafted messages, which are both time-consuming to create and must be updated whenever new features are introduced, then LLMs could fully automate their generation. 
This study considered explanations generated by an open-source model (Llama 3.3) and a closed-source commercial model (Claude 3.5 Sonnet).  These two LLMs were considered to evaluate whether a commercially available LLM (Claude) could provide better explanations than an open-weight solution such as LLama 3.3. This aspect is critical, as an organization may have strict privacy requirements that prohibit the sharing of sensitive data (e.g., emails) with third parties, such as Claude \citep{Shashidhar2023LLMsOpenSource,Panwar2024ChatgptLlama,Zhang2024LLMsMedicalOpenCommercial,Adams2024LlamaProprietaryRadiology,Espejel2025LLMs}. 

Building on these premises, we aim to address the following research questions:
\begin{itemize}
    \item \textbf{RQ1} Can LLMs generate explanations as effective as manually-generated ones in protecting users in warning dialogues? 
    
    \item \textbf{RQ2} Are feature-based explanations as effective as counterfactual explanations in warning dialogues?
    
    \item \textbf{RQ3} How do users perceive LLM-generated warnings?
    
    \item \textbf{RQ4}: How do user-related factors affect click-through rate (CTR) of phishing links?
\end{itemize}

In the following subsections, we detail i) how we designed the warning dialogues, ii) how the two explanation styles have been integrated into the warning dialogues, and iii) the selection process of the two LLMs considered in this study.

\subsection{Warning dialogues design}
\begin{figure*}[t]
    \centering
    \includegraphics[width=0.75\textwidth]{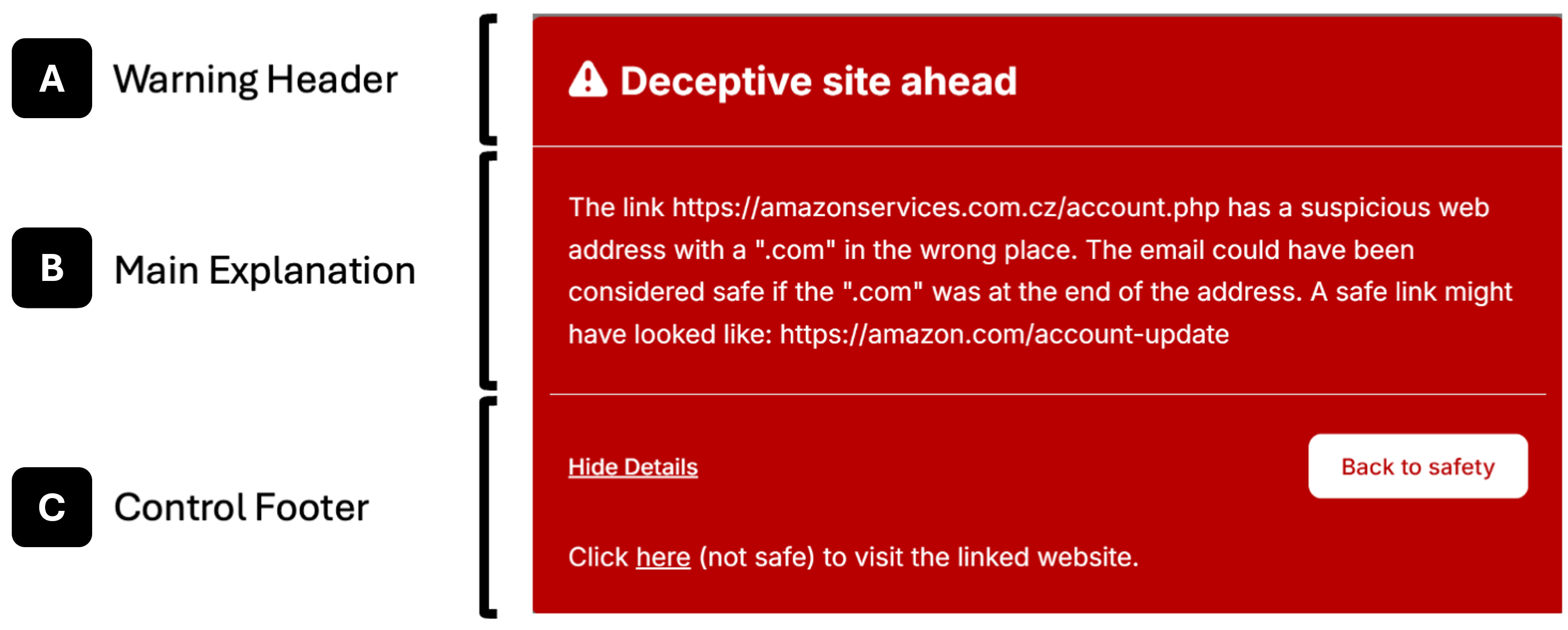}
    \caption{Anatomy of the warning dialogue used in the study. The top part (A) includes a brief warning that serves as the header of the box. The main part (B) presents the explanation message, while the footer (C) displays controls to return to the email visualization or proceed with browsing.}
    \label{fig:warning_structure}
\end{figure*}

The phishing warning dialogues explored in this study have the same structure to isolate the effects of the explanation message, depicted in Figure~\ref{fig:warning_structure}. 
A dialogue opens with an attention-grabbing header flagging the link as potentially dangerous, followed by a summary that attributes this judgment to the AI system. The main part of the interface is the explanation message, presented as a short paragraph embedded within the dialogue box. The bottom part contains buttons to get back to the email visualization or, with a less prominent presentation, a link to continue browsing the destination. All messages adhere to the same typographic style and layout to minimize visual biases. This design ensured that any differences in user response could be attributed to the nature of the explanations rather than variations in formatting.

\subsection{Explanation Styles}
The explanation message included in the main part differs according to the considered styles. Feature-based explanations emphasize observable cues extracted from the email, such as sender identity mismatches or suspicious links. The left part of Figure~\ref{fig:llms-warnings} shows sample explanations for this style, realized through declarative statements that point to concrete phishing indicators.  All messages share a common structure consisting of three sentences, inspired by prior work~\cite {desolda2023explanations}, which builds on the Communication-Human Information Processing (C-HIP) model \citep{wogalter2018communication} and previous studies on effective warning communication \citep{bauer2013warning}. The same structure was explicitly adopted when prompting the LLMs to generate the warnings, ensuring consistency across human- and AI-authored messages.

In contrast, counterfactual explanations present a hypothetical scenario in which the email would not have been flagged. These explanations reframe the AI's decision as contingent on a minor change, inviting users to consider alternative options. The examples adopt a conditional format, underscoring a key difference in framing: while feature-based explanations justify the current decision, counterfactuals highlight what could have made the outcome different. The right part of Figure~\ref{fig:llms-warnings} shows sample warnings using a counterfactual style.

\subsection{Generating the explanations}
\begin{figure*}[ht]
    \centering
    \includegraphics[width=0.75\textwidth]{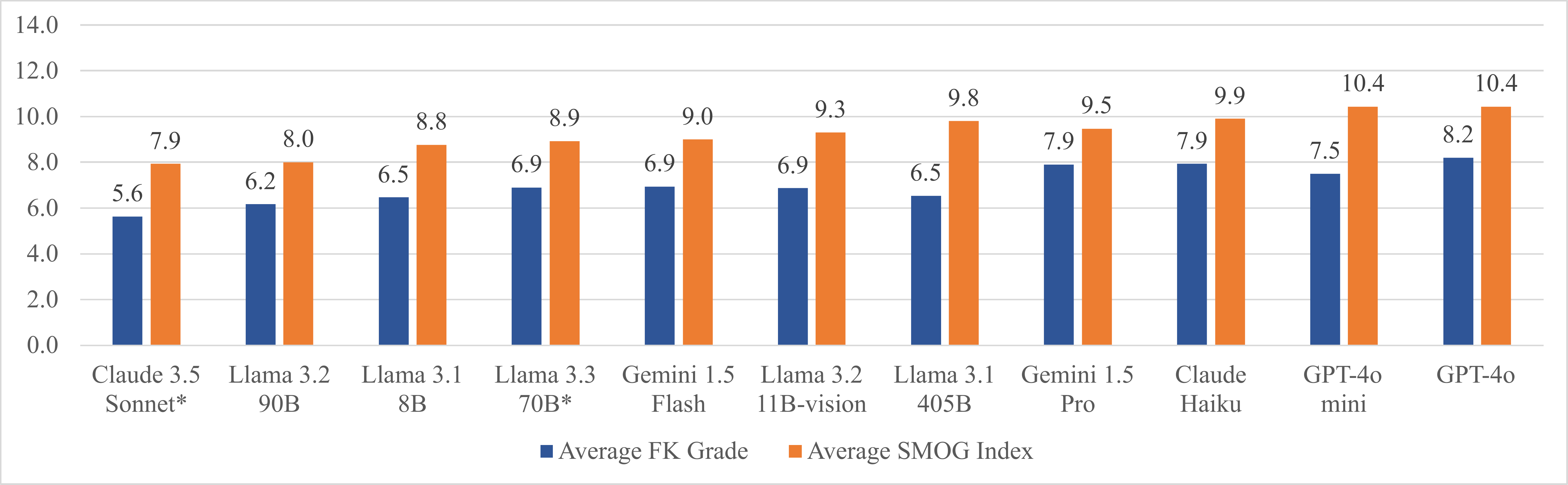}
    \caption{Average understandability and readability metrics for the feature-based explanations generated by the different LLMs — the lower, the better. The asterisks indicate the LLMs that were finally chosen.}
    \label{fig:fb_read_metrics}
\end{figure*}

\begin{figure*}[ht]
    \centering
    \includegraphics[width=0.75\textwidth]{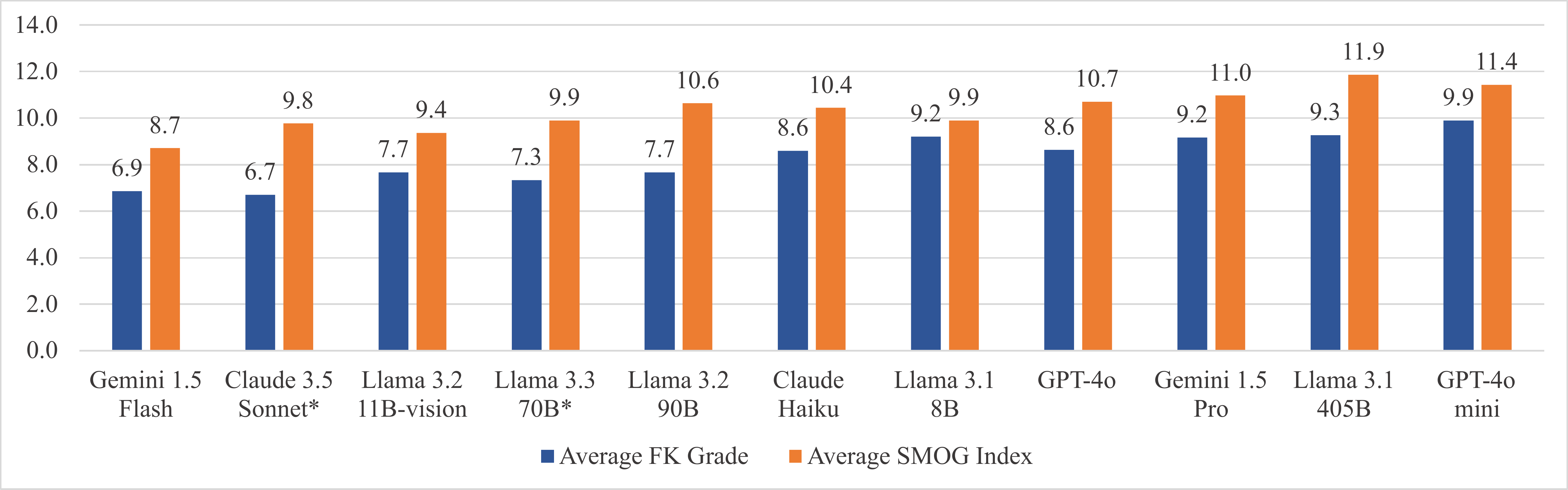}
    \caption{Average understandability and readability metrics for the counterfactual explanations generated by the different LLMs — the lower, the better. The asterisks indicate the LLMs that were finally chosen.}
    \label{fig:cf_read_metrics}
\end{figure*}

The other difference in the experimental conditions is the LLM used to generate the warning text. We selected the two LLMs to create explanations for this study through a rigorous evaluation process. Notably, 13 different LLMs (six commercially available and seven open-source) were included in our evaluation process as potential candidates. As grounds for this evaluation, we considered the three emails (two true positives and one false positive) whose links, when clicked, generated a warning dialogue during the experiment.
To evaluate each LLM, we followed this process:
\begin{itemize}
    \item Following the approach and prompts proposed in \cite{desolda2025apollo}, for each of the three emails, the LLM was prompted to generate an explanation message to be included in the warning, given the email body and one of the features to be explained. We considered the following three email features, which were found to be helpful to users when making decisions regarding phishing content \citep{greco2025enhancing, desolda2025apollo}, i.e.,
    (i) the Top-Level Domain of the URL in the email is misplaced (i.e., ``www.amazon.com.cz''); (ii) the URL in the email is an IP address; (iii) the link shown in the email and its actual destination mismatch (e.g., "click here" is shown instead of "www.facebook.com/...") --- this feature was found in the false positive email. The full prompts, extracted from~\cite{desolda2025apollo}, are reported in Appendix~\ref{appendix:prompts}.
    \item For each of the three emails, we generated a counterfactual explanation message through a modified version of the prompt proposed in \cite{desolda2025apollo}. This prompt is also reported in Appendix~\ref{appendix:prompts}.  
    \item Two researchers qualitatively assessed the six explanation messages to highlight whether an explanation was not generated correctly (e.g., by reaching the token limit early) or contained hallucinations. In this case, the previous step was repeated, and the produced explanation was analysed again for a maximum of three times. If every attempt did not lead to a coherent explanation, the LLM was marked as inadequate for this task. This process led to the discarding of Llama 3.2 1B-Instruct.
    \item The explanations underwent an automatic measurement of text readability with the FK (Flesch-Kincaid) Grade Level \citep{kincaid1975derivation} and of text understandability with the SMOG (Simple Measure of Gobbledygook) Index \citep{McLaughlin1969SMOG}.
    \item Explanations with low SMOG Index and FK Grade values were preferred as they indicate more readable and understandable text. Feature-based explanations' readability metrics are shown in Figure \ref{fig:fb_read_metrics}, while counterfactual-related ones are shown in Figure \ref{fig:cf_read_metrics}.  
\end{itemize}

To select the best commercial and open-weight LLMs, we initially selected a set of LLMs with the lowest, thus best, SMOG Index and FK Grade for both feature-based and counterfactual explanations. From this analysis, we selected Claude 3.5 Sonnet and Gemini 1.5 Flash as the best candidates for commercial LLMs, while all the Llama LLMs are candidates for the open-weight solutions. Then, two experts in HCI and cybersecurity performed a manual inspection of the generated explanations. This analysis removed Gemini 1.5 Flash because it generated explanations that were either too generic or contained hallucinations, especially for the counterfactual style (e.g., for the Amazon phishing email, it generated an explanation reporting ``[..] \textit{A safe email address wouldn't have strange parts in its name.} [..]''). Moreover, Llama 3.2 90B (as well as Gemini 1.5 Pro) was excluded, as it could not be forced to produce an explanation for the false-positive email. Llama 3.1 8B was a good candidate for feature-based explanations, although readability and understandability metrics were mediocre in the context of counterfactual explanations. Moreover, the generated feature-based explanation for the Amazon email contained a hallucination (``[..]\textit{ It looks like an Amazon link, but the '.com' name is typically found in other types of companies.} [..]''). Therefore, considering the qualitative assessment and the average metrics for both explanation styles, we selected Llama 3.3 70B as it represents a good trade-off between the resources needed to deploy this LLM and the quality of the explanations. 
The explanations produced by Claude 3.5 Sonnet and Llama 3.3 70B were then featured in the user study as described in Section \ref{subsec:study_design}.

It is worth noting that while our companion work APOLLO~\citep{desolda2025apollo} validates the technical scalability of generating explanations for thousands of emails, this study adopts a \textit{Human-in-the-Loop} validation approach for the experimental stimuli. We manually verified the LLM outputs, not because the models cannot generate text at scale, but to ensure that the experimental stimuli presented to the N=750 participants were free of hallucinations that could invalidate the internal validity of the user study. Thus, we validate the \textit{potential} of high-quality automated explanations.

\subsection{Claude vs Llama} 
While selecting the two LLMs to be used in the study, we identified some general characteristics of the generated explanation text. 

Claude tends to produce detailed and elaborate explanations. Its feature-based explanations frequently embed causal logic and downstream outcomes: for instance, a warning is issued that supplying data on a spoofed website risks account takeover. Within counterfactual scenarios, Claude employs conditional phrasing (“The email would have been safe if…”) while offering idealized reference links illustrating what a genuine message might look like. This is a continually organized discussion that incorporates technical pointers and a clear emphasis on the user experience that may occur.

\begin{figure*}[ht]
    \centering
    \includegraphics[width=0.75\textwidth]{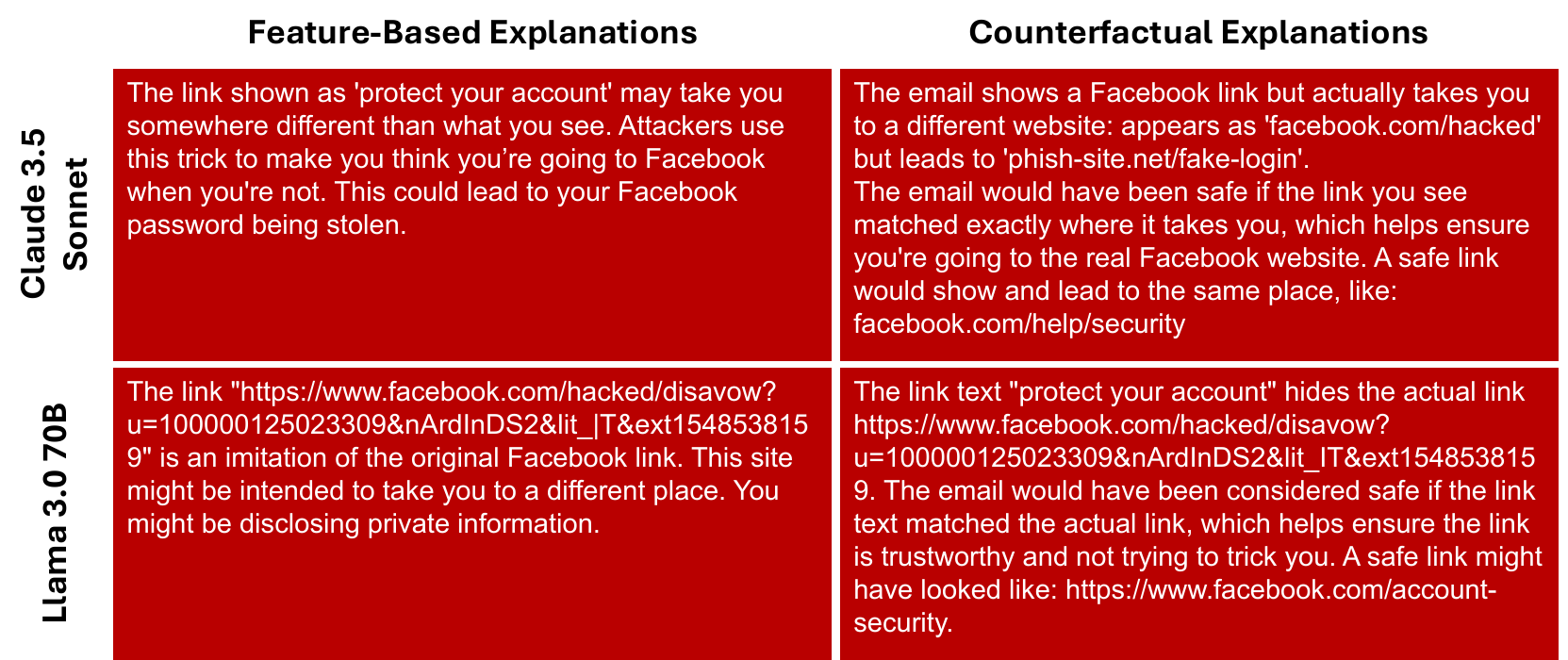}
    \caption{Four versions of the warning dialogue for the same deceptive link. Rows correspond to the two LLMs used in this study (Claude 3.5 Sonnet and Llama 3.3 70B), while columns correspond to explanation styles (feature-based and counterfactual).}
    \label{fig:llms-warnings}
\end{figure*}

Instead, Llama delivers explanations that are somewhat more concise and technical in tone. Its feature-based outputs highlight specific anomalies (e.g., numeric IP addresses, suspicious domain extensions) with minimal elaboration. The counterfactuals follow a similar structure to Claude’s but are generally more direct and less descriptive, focusing on the change required to alter the classification without extending into consequences. Llama also tends to use simpler language and more neutral phrasing, which may be more mechanical but still effective.
These stylistic differences are illustrated in Figure~\ref{fig:llms-warnings}. The top row includes examples generated with Claude 3.5 Sonnet, while the bottom row includes examples generated with Llama 3.3 70B. 

\section{Methods} \label{sec:methods}
In this section, we present a controlled experiment that compares the effectiveness of various explanations displayed in warning dialogues to protect users from phishing emails. The study specifically focused on evaluating explanations based on two factors we identified and discussed in the previous section: the \textit{explanation style} (either feature-based or counterfactual) and the \textit{LLM model} used to generate the explanation message (LLama 3.3 or Claude 3.5 Sonnet).
We also considered a baseline warning that included manually generated messages, which was empirically demonstrated to outperform all existing warnings for phishing attacks \citep{greco2025enhancing}. The study procedure and materials (i.e., warning design and emails included) were identical to those in \citet{greco2025enhancing} to ensure a fair and rigorous comparison with the baseline.  

\subsection{Study Design} \label{subsec:study_design}
This study employed a between-subjects design, with the explanation style and LLM used to generate the explanations as the independent variables. 
Five total between-subject levels were considered, four experimental warnings plus a baseline:

\begin{enumerate}
    \item \textbf{Llama Feature-based}: warnings containing feature-based explanations generated by the open-weight LLM Llama 3.3 70B.
    \item \textbf{Llama Counterfactual}: warnings containing counterfactual explanations generated by the open-weight LLM Llama 3.3 70B.
    \item \textbf{Claude Feature-based}: warnings containing feature-based explanations generated by the commercial LLM Claude 3.5 Sonnet.
    \item \textbf{Claude Counterfactual}: warnings containing counterfactual explanations generated by the commercial LLM Claude 3.5 Sonnet.
    \item \textbf{Baseline}: warnings that included feature-based explanations, which were manually written by experts (existing dataset - from \cite{greco2025enhancing}) 
\end{enumerate}

To improve the external validity of the study, for each experimental condition, we generated three explanations related to three different features. The complete list of explanation messages contained in each warning is reported in Appendix~\ref{appendix:warnings}. 

We excluded a "no-explanation" as a baseline since previous research has explicitly established its inferiority. Notably, \citet{greco2025enhancing} demonstrated that feature-based explanations significantly reduce phishing CTR (from 20.8\% to 11.8\%, $p=.032$) compared to generic warnings. Similarly, \citet{buono2023warnings} validated the superiority of active, explanatory warnings over passive indicators. Consequently, our study focuses on benchmarking automated LLM generation against this expert-crafted "gold standard", rather than re-validating the known baseline efficacy.

\subsection{Participants}
A total of 750 participants were involved, with equal allocation (n = 150) to each condition. The manually generated feature-based explanation condition included existing data (collected separately, n = 150), while the remaining 600 participants were recruited specifically for this study through the Prolific online platform\footnote{\url{https://www.prolific.co}}. The inclusion criteria for participants were as follows: i) fluency in English, ii) an even split between men and women (50-50), and iii) 18 or more years of age. A total of 61 participants either failed attention checks or were identified as inattentive and were thus excluded. To balance participation across the experimental conditions, we recruited an additional 61 participants to reach a final sample of 750 participants. The final sample had an average age of 32.98 years (sd = 11.01) and reported spending 7.94 hours per day on the Internet (sd = 3.95). 
The participants consisted of 375 males and 375 females. The participation in the study lasted approximately 20 minutes, and participants were rewarded with \pounds3.00, in line with Prolific's recommended participation fee of \pounds9.00/hour.

\subsection{Materials}

We developed a web platform with Laravel 9\footnote{\url{https://github.com/IVU-Laboratory/llm_warnings_explantions}} to conduct the study online. A total of 14 emails were included in the study: 2 of them were (harmless) phishing emails, while the remaining 12 were genuine ones designed starting from legitimate services/companies. 

We purposely employed a limited set of phishing scenarios (2 phishing emails out of 14) to maintain a realistic prevalence ratio (~14\%) without extending the session duration beyond 20 minutes, which would induce fatigue effects. While this limits the generalizability across all possible phishing vectors, it ensures high ecological validity regarding user attention spans in a simulated work environment. Future work should adopt longitudinal designs to test a broader array of threats.

A warning dialogue, such as that shown in Figure \ref{fig:warning_structure}, was shown to users who clicked on a phishing link. 
To simulate a case in which the system mistakenly classifies a genuine email as phishing, i.e., a false positive situation, we forced the platform to display a warning for one of the genuine emails as well. After the study, each user may have seen a maximum of three warning dialogues.

A full suite of user interactions was recorded in this study, including hovering over phishing links, clicking on them, visiting associated URLs, and responding to warnings, which enabled the calculation of the click-through rate (CTR) for each participant. The CTR was the proportion of accessed suspicious links to the number of displayed warnings, serving as an objective measure of the warning's effectiveness.

The survey instrument employed by 
\cite{desolda2023explanations} was administered to comprehend users' perceptions of the warnings. All the questions and their formats of responses are given in Table \ref{tab:warning_questionnaire}.

To assess the effects of users' knowledge of basic cybersecurity concepts, a 10-item test questionnaire was employed \citep{olmstead2017securityquestionnaire}.

Finally, a short, 6-item version of the Need for Cognition Scale (NCS-6) \citep{coelho2020needforcognition6item} was used to measure users' \textit{Need for Cognition}, which is the tendency to engage in and enjoy effortful cognitive activities \citep{cacioppo1982needforcognition}. Previous work highlighted that individual differences like NFC might significantly impact the effectiveness of AI support and explanations in terms of performance, overreliance on AI, and learning \citep{Cacioppo1996NFCCognitive,Bucinca2021NFC,Gajos2022NFC,Bucinca2024Optimizing}

\definecolor{Gray}{gray}{0.95} 

\begin{table}
  \caption{Questionnaire used to evaluate warning dialogues}
  \label{tab:warning_questionnaire}

  \rowcolors{2}{Gray}{white}
  \begin{tabularx}{\linewidth}{c X X}
    \toprule
    \textbf{\#} & \textbf{Question} & \textbf{Possible answers} \\
    \midrule
    1 & Did you read the entire text of the warning dialogue? & [yes; partially; no] \\
    2 & When you saw the warning dialogue, what was your first reaction? & [free text] \\
    3 & I understood the warning dialogue. & [5-point Likert: Strongly disagree – Strongly agree] \\
    4 & I am familiar with this warning dialogue. & [5-point Likert: Strongly disagree – Strongly agree] \\
    5 & I am not interested in this warning dialogue. & [5-point Likert: Strongly disagree–Strongly agree] \\
    6 & Which word(s) did you find confusing or too technical? & [free text] \\
    7 & Please rate the extent of risk you feel you were warned about. & [very low; low; no risk; risky; very high] \\
    8 & What action, if any, did the warning dialogue want you to take? & [continue; be careful while continuing; do not continue; nothing] \\
    9 & What do you think this warning dialogue means? & [free text] \\
    10 & Please rate your level of trust in this warning dialogue. & [not at all confident; not very confident; neutral; confident; very confident] \\
    11 & What is the first word in this warning dialogue? & [free text] \\
    \bottomrule
  \end{tabularx}
\end{table}

\subsection{Procedure}

\begin{figure}[ht]
    \centering
    \includegraphics[width=\linewidth]{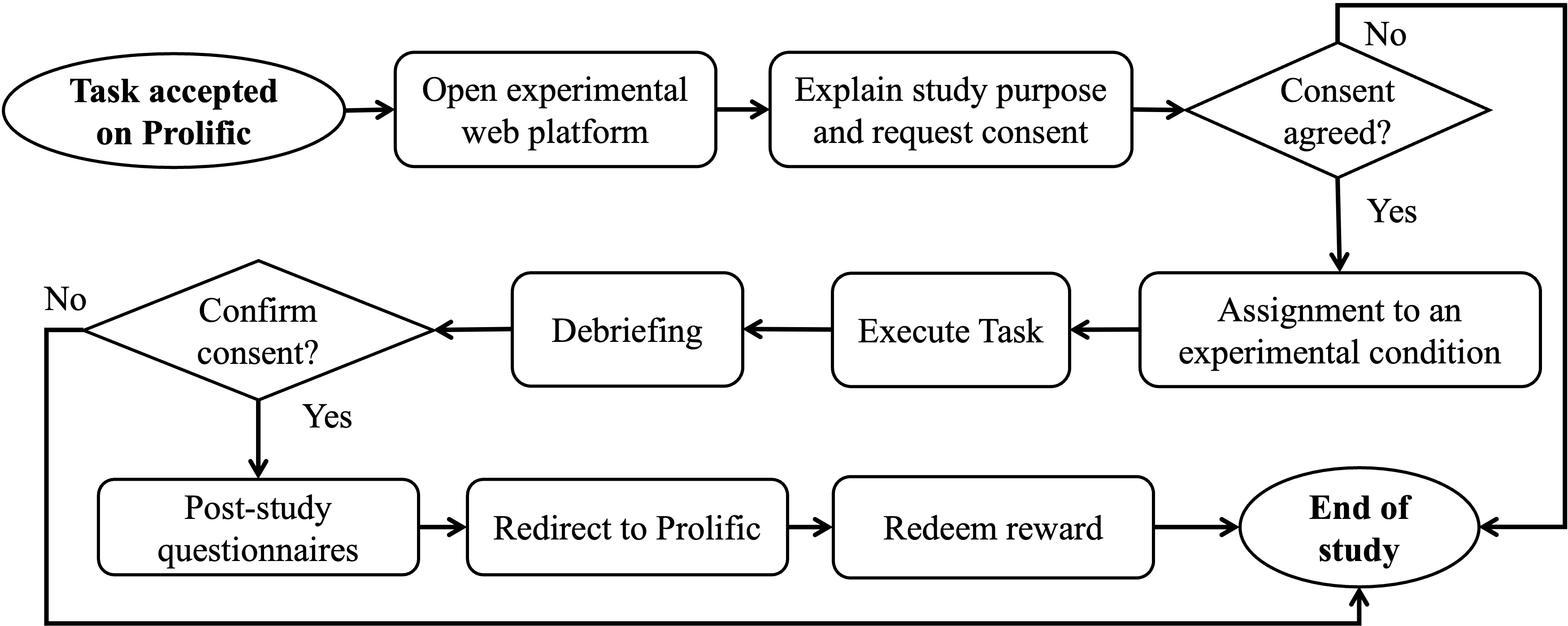}
    \caption{Study procedure (adapted from \cite{greco2025enhancing}).}
    \label{fig:study_procedure}
\end{figure}

The study procedure followed the same one in \cite{greco2025enhancing} and is depicted in Figure \ref{fig:study_procedure}.
To prevent priming effects regarding the experiment’s true objective and to mitigate the \textit{Hawthorne effect} \cite{macefield2007usability}, participants were initially deceived about the study’s real purpose ~\citep{sotirakopoulos2011challenges}. Deception is one of the most adopted techniques in cybersecurity usability studies \cite{distler2021slr}. In our context, explicitly mentioning "phishing" or "security" would have biased participants towards a suspicious mindset, unrepresentative of daily email usage. Therefore, participants were told the aim was to evaluate the user experience of a new email client interface. This framing encouraged participants to focus on the functionality and usability of the client, simulating a realistic scenario of split attention where security is a secondary task.

After the Prolific task was accepted, the participants were redirected to a web platform where the study was hosted. The (deceptive) purpose of the study, the data usage policies, and the withdrawal right were described on the landing page. 
According to the ethical guidelines, participants were asked to provide their informed digital consent. All data on participants and interactions were completely anonymized and securely stored on the university servers. Any concerns that the participants may have had could be addressed after the study. The design was such that it did not cause excessive psychological stress; for example, phishing scenarios were realistic yet not too alarming.

Once consent was obtained, all participants were randomly assigned to four experimental conditions, with balanced participation across conditions. Then, participants were asked to follow a case study of a fictional user named Alice, who was testing a new email client she had adopted at her workplace. The respondents were asked to interact with the email client by reading messages in the inbox and trying the functionality of embedded links.

\begin{figure}[ht]
    \centering
    \includegraphics[width=\linewidth]{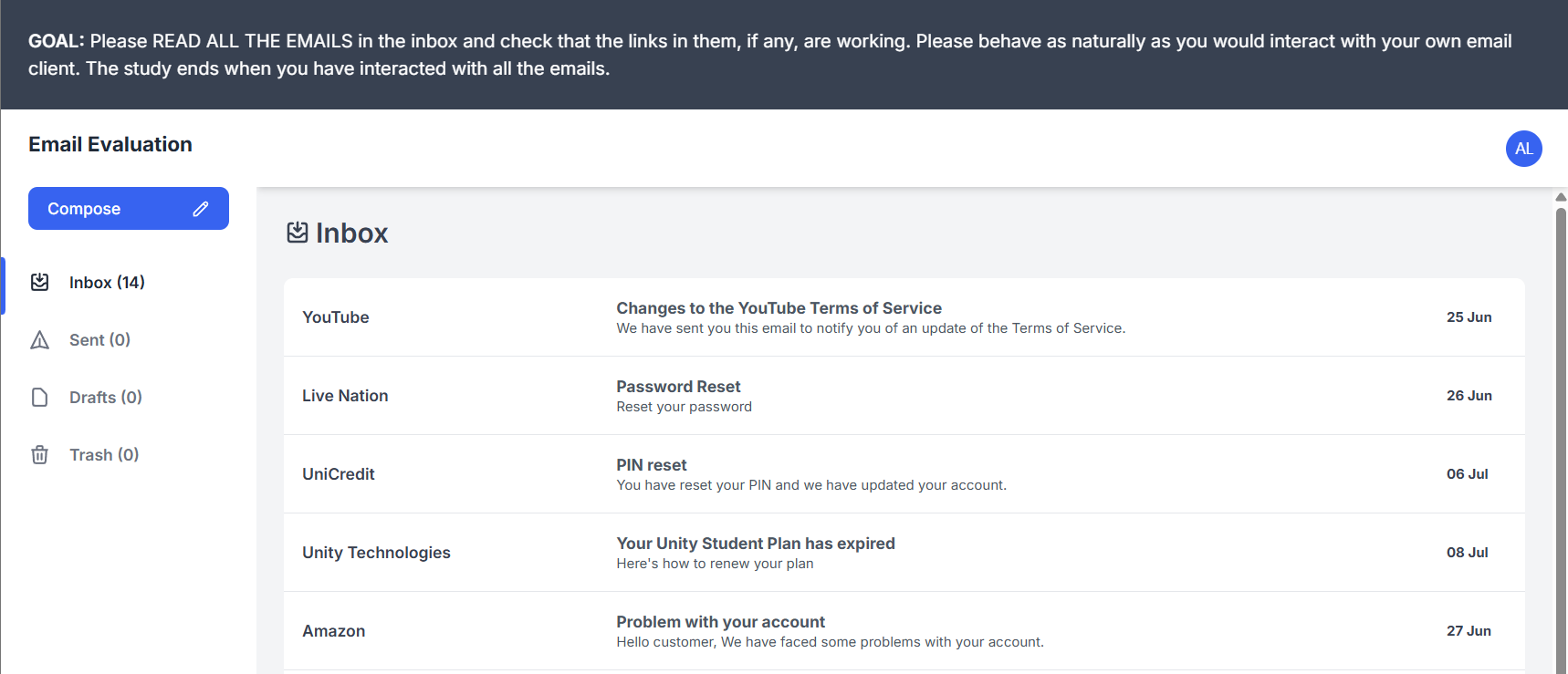}
    \caption{Web platform used in the study.}
    \label{fig:study_platform}
\end{figure}

\begin{figure}[ht]
    \centering
    \includegraphics[width=\linewidth]{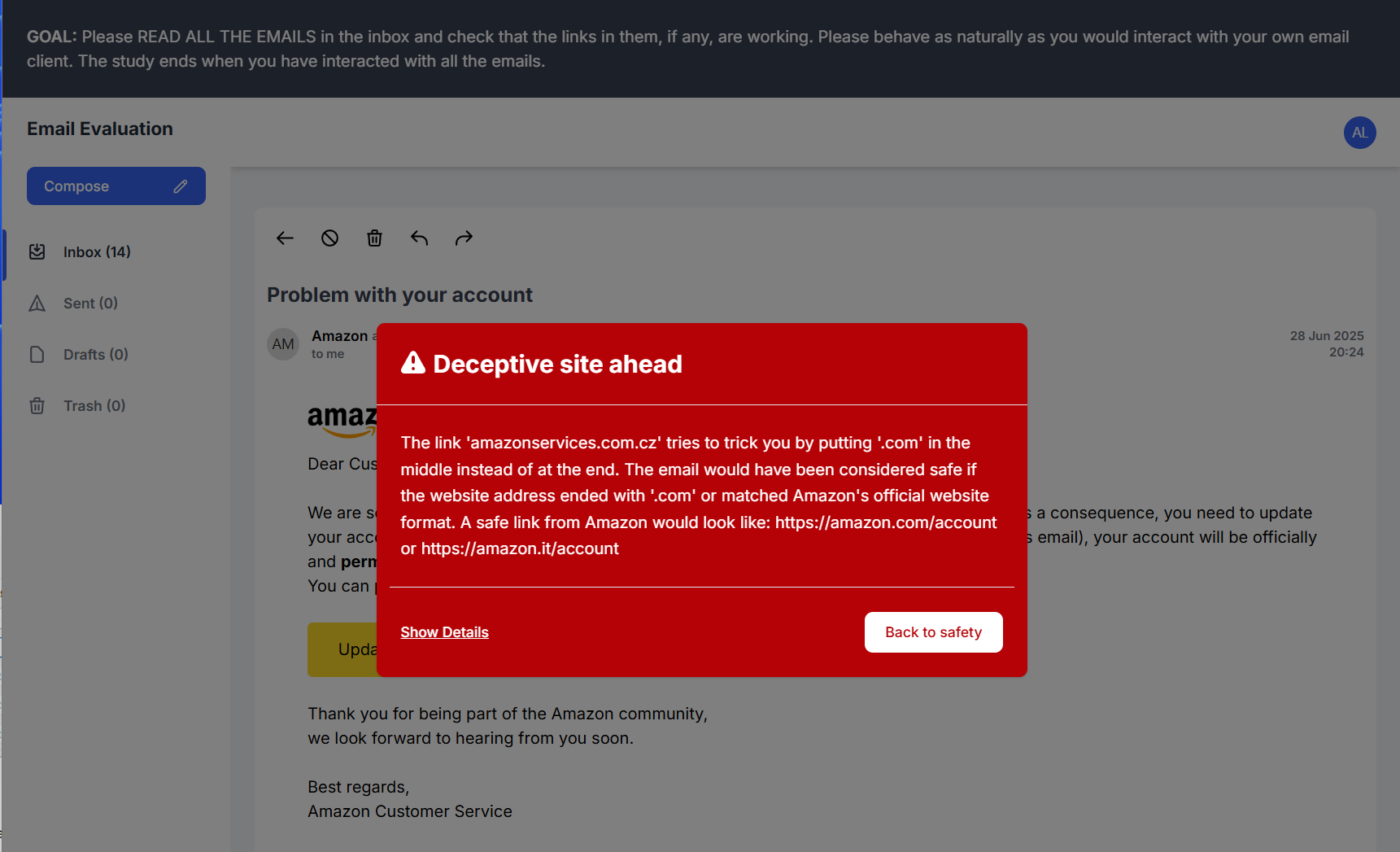}
    \caption{Warning shown during the study for a fake Amazon email containing a mimicking URL with a suspicious domain (``.com.cz'').}
    \label{fig:study_platform_warning}
\end{figure}

The scenario was presented in a banner above the simulated email client interface, which displayed randomised email previews, including the sender's name, title, and subject (Figure \ref{fig:study_platform}). Clicking an email revealed its content; phishing links triggered warnings (Figure \ref{fig:study_platform_warning}). After each interaction with a phishing message (via link click or by navigating back), participants answered two attention-check questions: \enquote{Who was the sender of the email?} and \enquote{Were there links in the email?}. These checks help identify inattentive respondents, which are especially typical in remote settings \citep{matsuura2021careless}.

Upon completing their interactions, participants were debriefed and informed of the actual goal of the study: assessing user responses to warnings triggered by clicks on suspicious links. The debriefing step is a critical ethical practice in studies involving deception \citep{sotirakopoulos2011challenges}. After the debriefing, participants were asked to reconfirm their consent. Those who chose not to continue would have had their data deleted; however, all participants opted to proceed. They then completed a series of questionnaires: first, the warning perception questionnaire from \cite{desolda2023explanations}, followed by the Raw NASA TLX workload assessment \citep{hart1986NASATLX}, a cybersecurity proficiency questionnaire \citep{olmstead2017securityquestionnaire}, and the Need for Cognition Scale \citep{coelho2020needforcognition6item}. Finally, participants completed a demographic survey that asked about their age, gender, and average daily internet use.
Participants were thanked for their involvement and redirected to Prolific to claim compensation. 

The study procedure was approved by the Research Ethics Committee of King's College London, reference number HR-23/24-41353.

\subsection{Data Analysis}

To answer RQ1 (\enquote{Can LLMs generate explanations that are as effective as manually-generated ones in protecting users in warning dialogues?}) and RQ2 (\enquote{Are feature-based explanations as effective as counterfactual explanations in warning dialogues?}), we applied the Chi-Square test of independence. 
When differences emerged, Bonferroni adjustment for pairwise comparisons was used as a post-hoc test to analyse differences among individual conditions.

For answering RQ3 (\enquote{How do users perceive LLM-generated warnings?}), we employed the Kruskal-Wallis test to analyse differences across the experimental conditions for items 3, 4, 5, 7, and 10 of the warning questionnaire (regarding \textit{understandability}, \textit{familiarity}, \textit{interest}, \textit{felt risk}, and \textit{trust}, respectively) and items of the NASA-TLX questionnaire. 
In the case of statistical significance, Dunn's test was applied to analyse differences between individual conditions.  
The Chi-Square test of independence was used to analyse item 8 of the warning questionnaire (relative to the intended action to take).

Moreover, to answer RQ3, a deductive thematic analysis (TA) was also conducted to examine qualitative responses from open-ended questions of the warning questionnaire (see items 2, 6, and 9 in Table \ref{tab:warning_questionnaire}) following Braun and Clarke’s six-step framework \citep{braun2006thematic}.

Finally, for RQ4 (\enquote{How user-related factors affect CTR?}), we employed a binomial logistic regression to analyse the effect of user-related factors, such as demographics, need for cognition, and warning perception, on CTR.

In each analysis of the RQs, three splits of the dataset were considered: firstly, the whole dataset to analyse the general CTR; secondly, only the data relative to the True Positive emails to examine the warning's level of protection against actual phishing emails; and thirdly, only the data relative to the False Positive email, to gain insights about the interaction with a warning dialogue that is wrong. 
Lower CTRs in the case of true positive emails indicate a higher level of protection, as the user is effectively avoiding actual phishing links. On the other hand, a lower CTR in the case of the false positive indicates that the warning does not support the user in recognising when the system is incorrect; thus, a higher CTR is preferable for the false positive condition only. All the quantitative data were analysed using R 4.2.2. An alpha level of .05 was used for all statistical tests.

\section{Results and Discussion} \label{sec:results}

In this section, we report the results of the controlled experiment to answer each research question. The results of the statistical tests are reported only when significant differences emerged; the full results of the statistical analysis are reported as additional resources in our \href{https://github.com/IVU-Laboratory/llm_warnings_explantions/blob/main/Submission%20-%20Additional%20Material/Results_additional_material.zip}{GitHub repository}.

\subsection{RQ1 - LLM vs Manual Explanations}
\begin{figure*}[ht]
    \centering
    \includegraphics[width=0.75\textwidth]{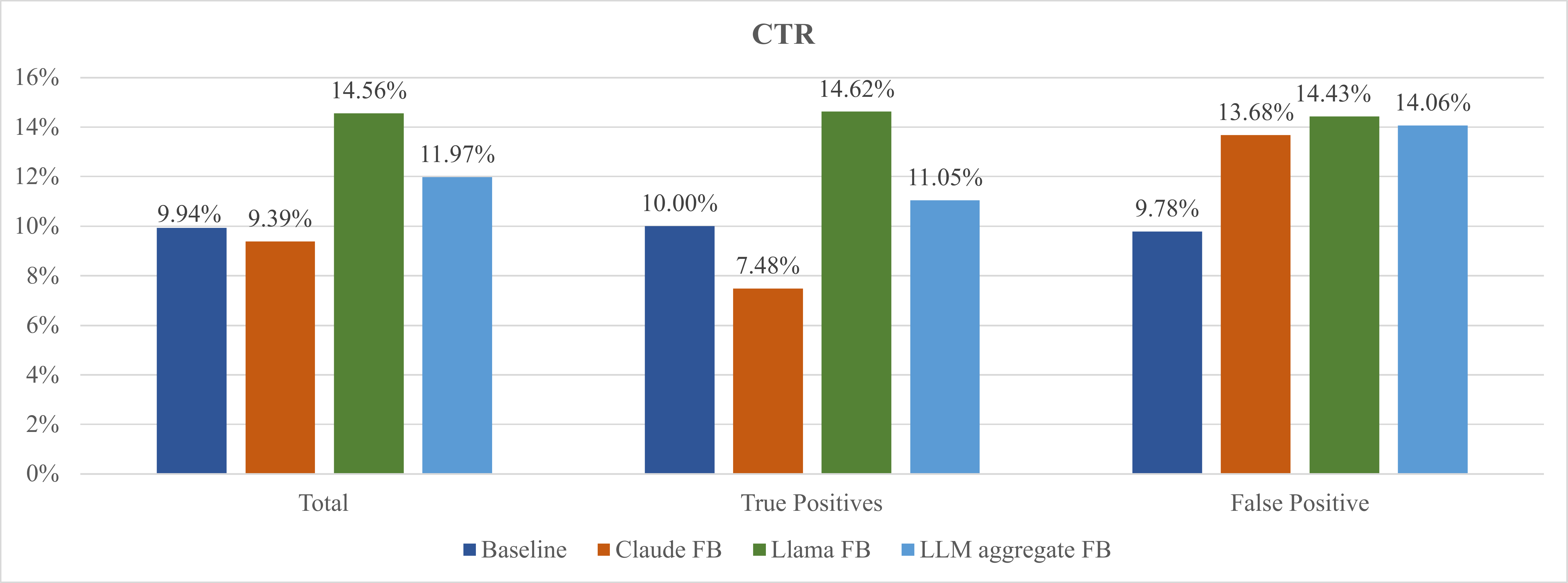}
    \caption{Click-through rates (CTRs) in the baseline condition (manually generated, feature-based explanation) and for LLM-generated feature-based explanations, shown individually (Claude FB, Llama FB) and aggregated.}
    \label{fig:rq1}
\end{figure*}

To answer the first research question, we compared the results of the experimental conditions with the baseline from \cite{greco2025enhancing}, which employed warnings with manually generated explanations. To obtain a fair comparison, we excluded from this analysis conditions that employed counterfactual explanations, as the baseline included a feature-based explanation (n = 300 for the aggregated experimental conditions, n = 150 for the baseline sample, totaling 450). The CTRs for the baseline and LLM feature-based conditions (both individual and aggregated) are reported in Figure \ref{fig:rq1}. 
As it can be observed, LLM-generated warnings generally led to a higher CTR when aggregated, compared to the baseline. A more in-depth analysis revealed that the Claude FB condition had the lowest CTR (9.39\%), especially for true positive emails (7.48\%), while the baseline led to the lowest CTR in the case of the false positives (9.78\%). 
On the contrary, Llama FB led to the highest CTR in all cases (14.65\% in total, 14.62\% for true positives, and 14.43\% for false positives). 

The difference in CTR between the Baseline and Claude FB was not statistically significant at the $\alpha = 0.05$ level ($p > 0.05$). This finding is fundamental: it demonstrates that replacing expensive human experts with an automated LLM (Claude) does not degrade user protection. The LLM's non-inferiority is a positive result for the scalability of phishing defense. However, comparing the baseline, Claude FB, and Llama FB led to a significant difference ($\chi^2 = 5.86$, df = 2, $p = 0.05$) in the number of True Positive emails. Although pairwise comparisons did not reveal differences, a trend emerged between Claude-generated feature-based warnings (8.90\%) and Llama-generated ones (14.29\%), as the post-hoc test indicates a marginal difference between the two ($p=0.084$).

\subsection{RQ2 - Feature-based vs Counterfactual Explanations }
\begin{figure*}[ht]
    \centering
    \includegraphics[width=0.75\textwidth]{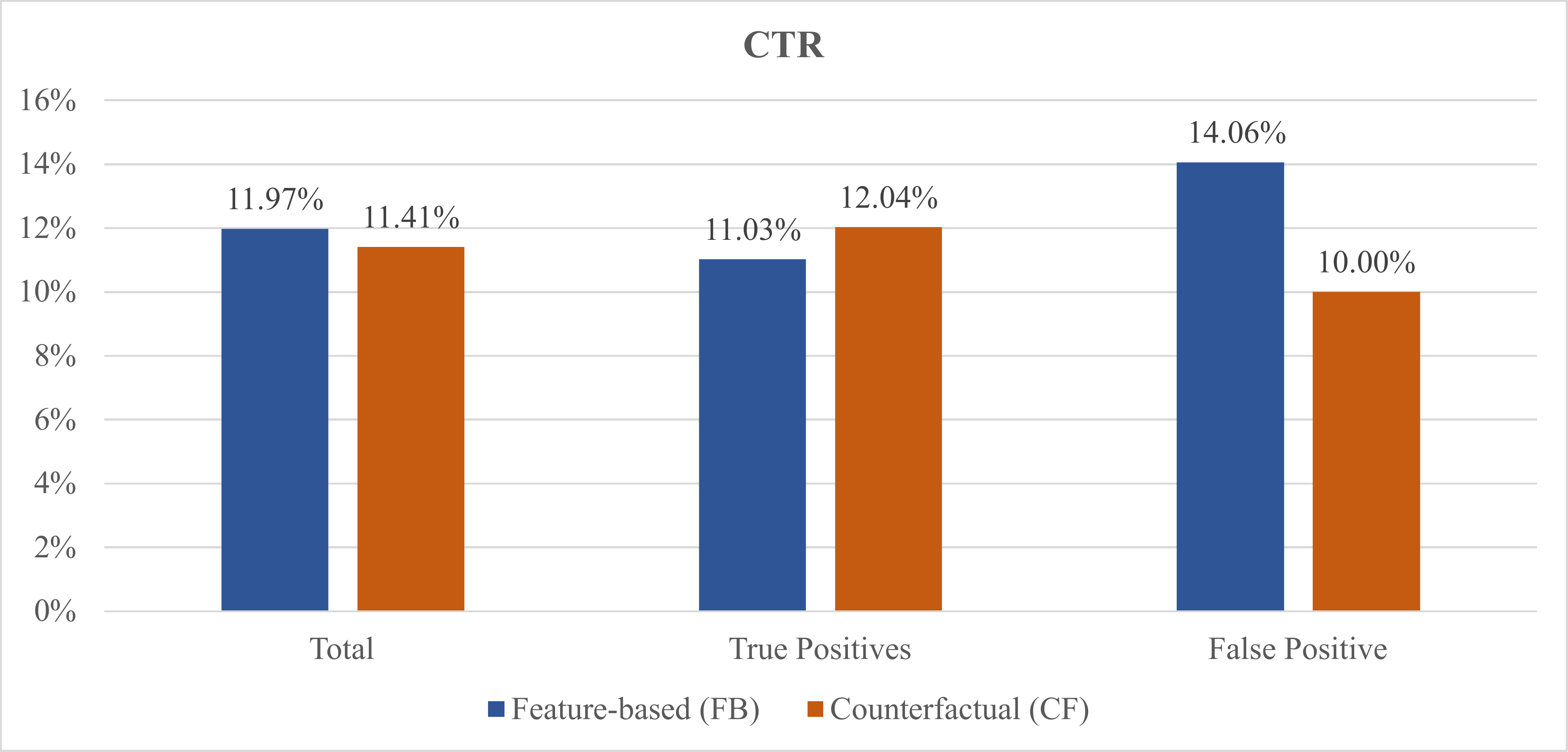}
    \caption{Click-through rates (CTRs) for aggregated conditions with feature-based and counterfactual explanations. The baseline condition is excluded.}
    \label{fig:rq2}
\end{figure*}

The second research question concerned the effectiveness of explanations in warning dialogues, specifically in terms of their type (feature-based or counterfactual). Therefore, we aggregated the conditions according to the explanation type they included and made a direct comparison (n = 300 for both samples, totaling n = 600). To have a balanced analysis, we excluded the baseline from this comparison. The CTRs for the conditions included in this analysis are reported in Figure \ref{fig:rq2}. 

Compared to the feature-based explanations, the warnings with counterfactual explanations led to a lower CTR in general (11.41\% vs 11.97\%) and for the false positive email (10.00\% vs 14.06\%), while slightly increasing the CTR for the true positives (12.04\% vs 11.03\%).

The Chi-Square test did not highlight any statistically significant difference between the two types of explanation, using the same LLM. 
This is true when either considering all the emails ($\chi^2 = 0.05$, df = 1, $p=0.827$), only the true positives ($\chi^2=0.12$, df = 1, $p=0.724$), or only the false positives ($\chi^2=1.13$, df = 1, $p=0.288$).

\begin{figure*}[ht]
    \centering
    \includegraphics[width=0.75\textwidth]{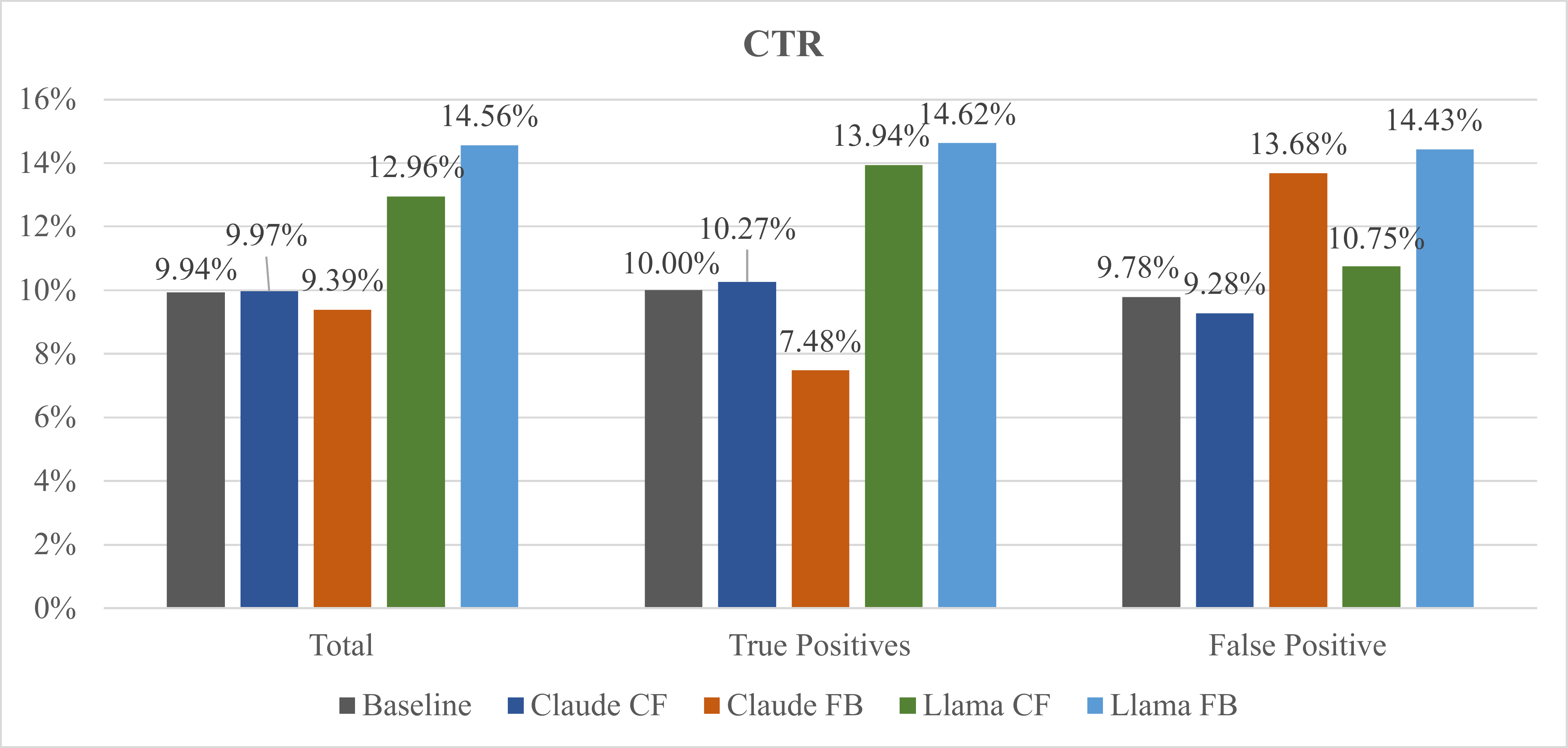}
    \caption{Click-through rates (CTRs) for individual experimental conditions and the baseline.}
    \label{fig:rq2b_ctr}
\end{figure*}

To deepen the analysis of RQ2, we investigated the CTR for the experimental conditions taken individually (n = 150 for each sample, totaling n = 600). Figure \ref{fig:rq2b_ctr} reports the results. 
It is noticeable how feature-based explanations from Claude led to the highest level of protection (CTR = 9.39\% in general and CTR = 7.48\% for true positives). In contrast, feature-based explanations from Llama yielded the highest CTR (14.56\% in general, 14.62\% for true positives, and 14.43\% for false positives). 
In the case of the false positive email, the baseline resulted in the lowest CTR (9.78\%). 

The Chi-Square test highlighted no significant differences. However, the true positive emails led to a slight statistical difference ($\chi^2 = 6.97$, df = 3, $p=0.073$). This result might suggest a difference in CTR between the Claude FB (7.48\%) and Llama FB (14.62\%) conditions.

\subsection{RQ3 - User Perception of Warnings}

The third research question explored how users perceive warnings with LLM-generated explanations. To achieve this, we examined the results of the warning questionnaire (see Table \ref{tab:warning_questionnaire}) and the raw NASA-TLX scores. Responses to the NASA-TLX were considered to measure the workload caused by the experimental condition. Moreover, to gain further insights, we also computed the six sub-dimensions of the NASA-TLX, namely mental demand, physical demand, temporal demand, perceived performance, perceived effort, and frustration level. 

\begin{figure*}[ht]
    \centering
    \includegraphics[width=0.75\textwidth]{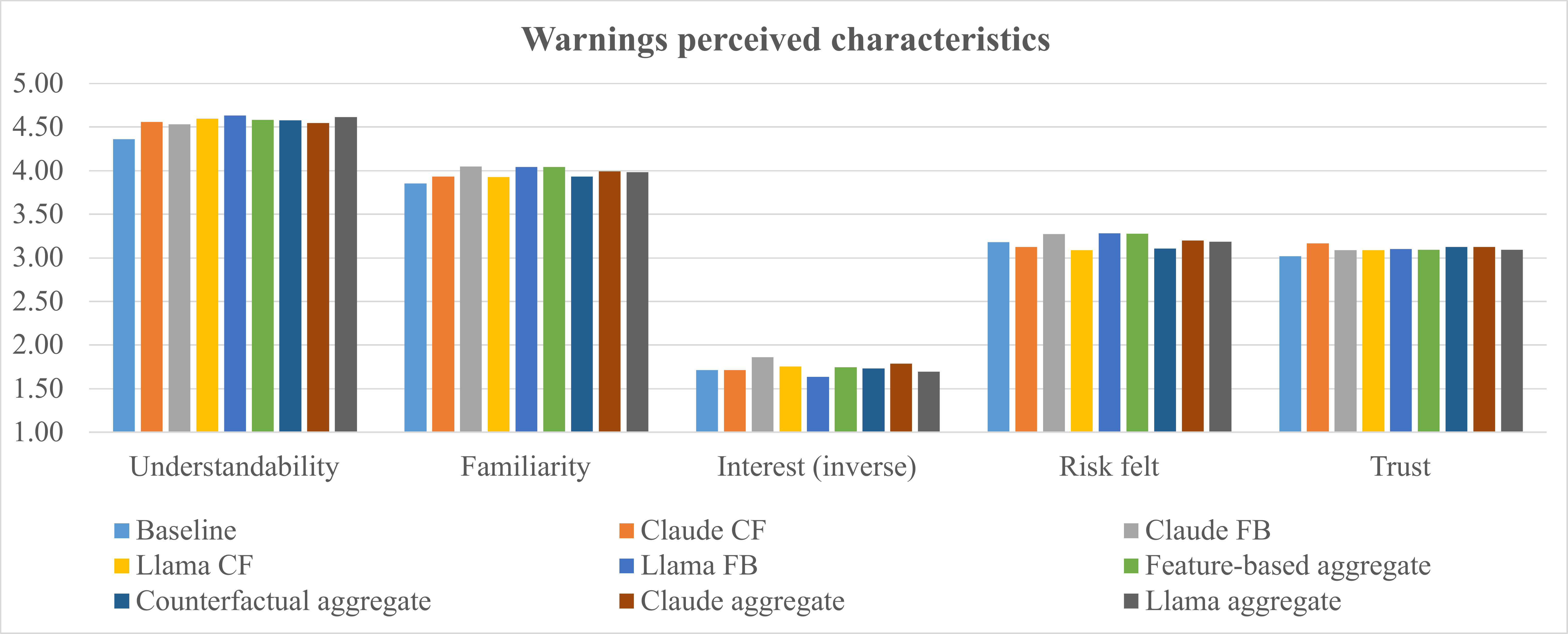}
    \caption{Average perceived characteristic values derived from the warning questionnaire results.}
    \label{fig:rq3_warning_perception}
\end{figure*}

\begin{figure*}[ht]
    \centering
    \includegraphics[width=0.75\textwidth]{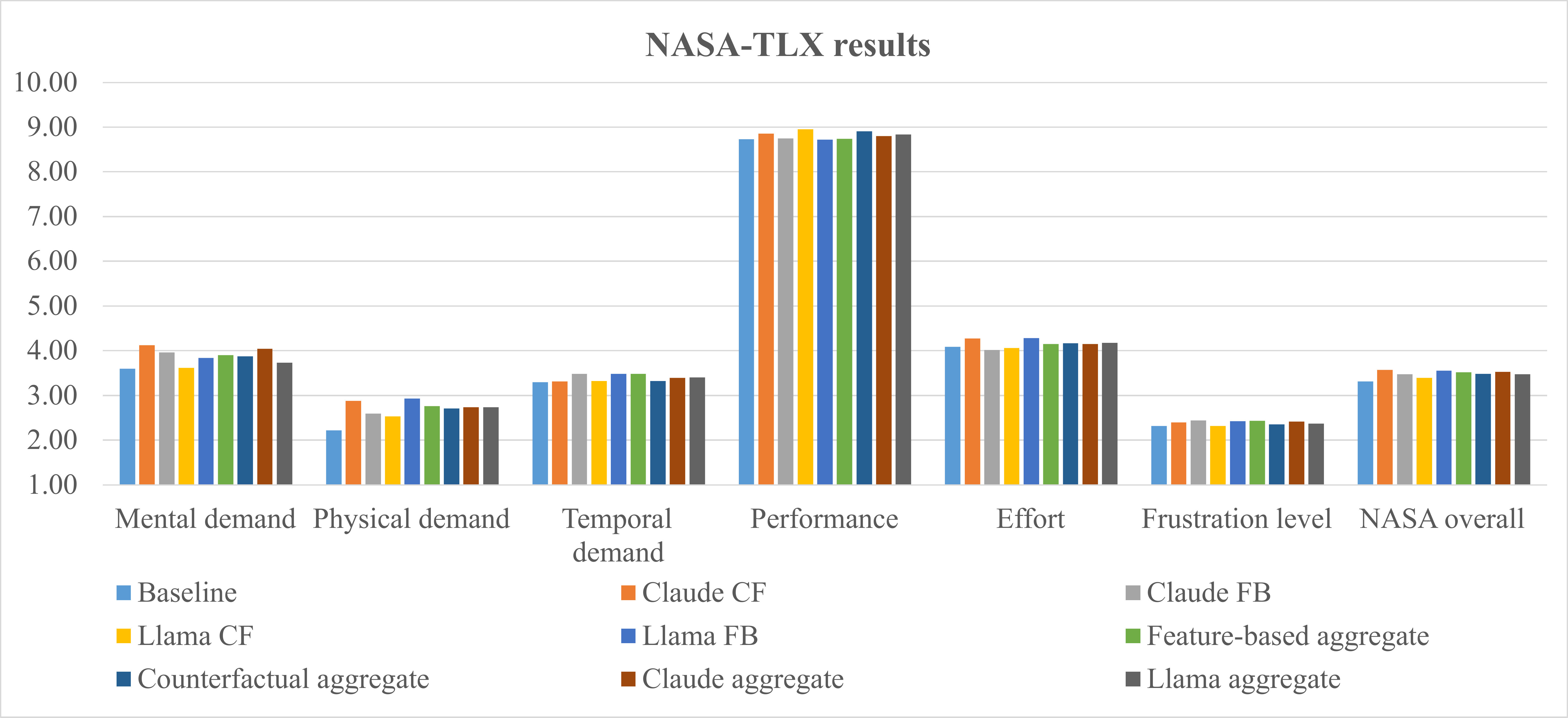}
    \caption{Average scores for each dimension of the Raw NASA-TLX and overall scores by condition and aggregation.}
    \label{fig:rq3_nasa}
\end{figure*}

The results of the extracted dimensions for the baseline, the four experimental conditions, and the aggregation based on explanation type and LLM are reported in Figures \ref{fig:rq3_warning_perception} and \ref{fig:rq3_nasa}.

From the Kruskal-Wallis test between the four experimental conditions and the baseline (n=150 each, n=750 in total), a statistically significant difference emerged for warnings' understandability ($H = 15.76$, df = 4, $p=0.003$). 
Dunn's post-hoc test revealed that warnings generated with Llama were perceived as significantly more understandable than the baseline, both in the case of feature-based ($Z= 3.53$, $p=0.002$) and counterfactual explanations ($Z = 3.29$, $p=0.005$).
No other significant differences emerged.

Item 8 of the warning questionnaire (\enquote{What action, if any, did the warning dialogue want you to take?}) required a separate analysis, as the data type was not numeric but categorical. The Chi-square test did not reveal any significant difference ($\chi^2=9.68$, df = 5, $p=0.644$).

\definecolor{Gray}{gray}{0.95} 

\begin{table}
\caption{Participants' actions (Item 8) by warning condition (\%).}
\label{tab:action_warnings}

\rowcolors{2}{Gray}{white}
\begin{tabular}{p{1.6cm} r r r r}
\toprule
\textbf{Condition} & \textbf{Don’t continue} & \textbf{Be careful} & \textbf{Continue} & \textbf{None} \\
\midrule
Baseline                  & 74.0\% & 24.0\% & 0.0\% & 2.0\% \\
Claude CF                 & 71.3\% & 25.3\% & 2.7\% & 0.7\% \\
Claude FB                 & 78.0\% & 18.7\% & 2.7\% & 0.7\% \\
Llama CF                  & 74.0\% & 21.3\% & 3.3\% & 1.3\% \\
Llama FB                  & 74.0\% & 24.0\% & 1.3\% & 0.7\% \\
FB agg.              & 76.0\% & 21.3\% & 2.0\% & 0.7\% \\
CF agg.              & 72.7\% & 23.3\% & 3.0\% & 1.0\% \\
Claude agg.          & 74.7\% & 22.0\% & 2.7\% & 0.7\% \\
Llama agg.           & 74.0\% & 22.7\% & 2.3\% & 1.0\% \\
\bottomrule
\end{tabular}
\end{table}

\subsection{RQ4 - How User-Related Factors Affect CTR}

To investigate how user-related factors affect CTR, we built a series of binomial logistic regression models that included demographic measures (age, gender, expertise, average hours online per day, and Need-for-Cognition) and warning-perception variables (understandability, familiarity, trust, perceived suggested action, and mental workload). Different models were trained for three outcomes: CTR across all emails (ALL), true positives (TP), and false positives (FP), across all conditions and LLM variants.

Moreover, odds ratios (OR) were computed by exponentiating the estimated coefficients from the logistic regression models, by applying the formula $\mathrm{OR} = e^{\beta}$. This transformation allows interpreting each $\beta$ coefficient as the multiplicative change in the odds of clicking associated with a one-unit increase in the predictor. An OR greater than 1 indicates a higher likelihood of clicking, whereas an OR below 1 indicates a lower probability. To further simplify the interpretation of the OR, we mapped the ORs into \textit{effect sizes}, categorised as small (OR 1.2–1.5 / 0.67–0.83), moderate (1.51–2 / 0.5–0.66), and large ($>$2 / $<$0.5), following the guidelines suggested in \cite{chen2010big}. The majority of significant effects observed in our data fell either into the "small" (OR~1.2–1.3) or "large" (OR$>$2 or $<$0.5) categories, with relatively few predictors yielding odds ratios in the moderate range.
Odds ratios of 0.84 to 1.19 were considered as negligible because they show a difference in odds of less than 20\% in clicking (which will practically be insignificant even in the case of a significant difference). This strategy is guided by the prescriptions of the applied behavioural research to differentiate between the statistical and practical significance of the requirements \citep{Kirk1996, fritz2012effect}.

\textbf{Familiarity} with warnings showed consistent large effects: higher familiarity was associated with markedly higher CTR in multiple conditions. ORs ranged from 1.8 to 3.24, with significant effects in Counterfactual aggregate (ALL, TP, FP), Claude Counterfactual (ALL, TP, FP), Claude aggregate TP, and Llama Counterfactual (ALL, TP).

\textbf{Interpreting the warning as suggesting “don’t continue”} led to large decreases in CTR across different conditions: ORs between 0.10 and 0.41, significant in Counterfactual aggregate (ALL, TP), NoLLM (ALL, TP), Llama aggregate (ALL, TP), and Claude aggregate (ALL, TP).

\textbf{Gender} showed large and asymmetric effects. Female participants consistently had much lower CTR: ORs as low as 0.0003 (NoLLM FP) and generally below 0.05 in several conditions (Counterfactual, NoLLM, Llama aggregate, Claude aggregate, Claude Counterfactual, Llama Counterfactual). Male participants had higher CTR in some conditions: OR=2.62 (Counterfactual aggregate TP), OR=2.13 (Llama aggregate TP), while in one case, male gender was associated with lower CTR (NoLLM FP, OR=0.13).

\textbf{NASA-TLX} (mental workload) was associated with moderate to significant reductions in CTR: ORs between 0.45 and 0.74, significant in Feature-based FP, Llama aggregate (ALL, TP), Counterfactual aggregate (ALL, TP), Claude aggregate (ALL, FP), Claude Feature-based FP, and Llama Feature-based FP.

\textbf{Understandability} also showed protective effects: higher scores correlated with lower CTR, with ORs between 0.48 and 0.51 across Counterfactual aggregate (ALL, TP) and Claude aggregate FP.

\textbf{Trust in warning} (Llama Feature-based FP) also showed a significant reduction in CTR (OR=0.44).

\textbf{Average hours online per day} had small positive effects: OR=1.24 (NoLLM ALL) and OR=1.22 (NoLLM TP).



Interactions between being \textbf{male} and \textbf{need-for-cognition} produced small to large effects: OR=1.44 in Counterfactual aggregate for false positives, OR=2.15 in Claude Counterfactual for false positives, OR=1.33 in Claude Counterfactual for all emails, OR=1.32 in Claude Counterfactual for true positives, and a smaller effect of OR=1.21 in Counterfactual aggregate for all emails. 




Other smaller effects were observed for the interaction between \textbf{expertise} \textbf{trust in warning} (OR=0.44). 

Full results, including ORs, p-values, and effect size categorisations, are detailed in Table~\ref{tab:user_factors_ctr} reported in Appendix~\ref{appendix:lr}.

\subsection{Qualitative results}
This subsection describes the themes identified through deductive thematic analysis of Items 2, 6, and 9, as presented in Table \ref{tab:warning_questionnaire}. This analysis strictly followed Braun and Clarke’s six-step framework \citep{braun2006thematic}: \textit{data familiarisation}, \textit{code generation}, \textit{theme identification}, \textit{theme review}, \textit{theme definition and naming}, and \textit{report production}. The deductive approach was selected to align the coding framework with the prior study that inspired our experimental design and provided the baseline warning condition \citep{greco2025enhancing}. 
Two independent researchers, with backgrounds in Human-Computer Interaction and qualitative research methods, carried out the process. They were blind to the participants' experimental conditions during the first phases to avoid bias. In particular, the researchers first familiarized themselves with the data by repeatedly reading transcripts and noting initial impressions. 
Then, they independently coded relevant excerpts. Inter-rater reliability was evaluated using Cohen’s kappa ($\kappa$) values: Item 2 = .81, Item 6 = .84, and Item 9 = .79. These values indicate good agreement \citep{landis1977application}. Discrepancies were resolved through discussion until a consensus was reached.   
Related codes were grouped into preliminary themes (80\% initial agreement), which were collaboratively refined to ensure an accurate representation of the data. Each theme was then clearly defined and named, and the final report included interpretive commentary and excerpts of supporting data. 
For each of the three items, the themes captured the essence and meaning of the users' answers, specifically their initial reactions to the warnings (item 2), terms perceived as confusing (item 6), and interpretations of warning meanings (item 9).

\begin{figure*}[ht]
    \centering
    \includegraphics[width=0.75\textwidth]{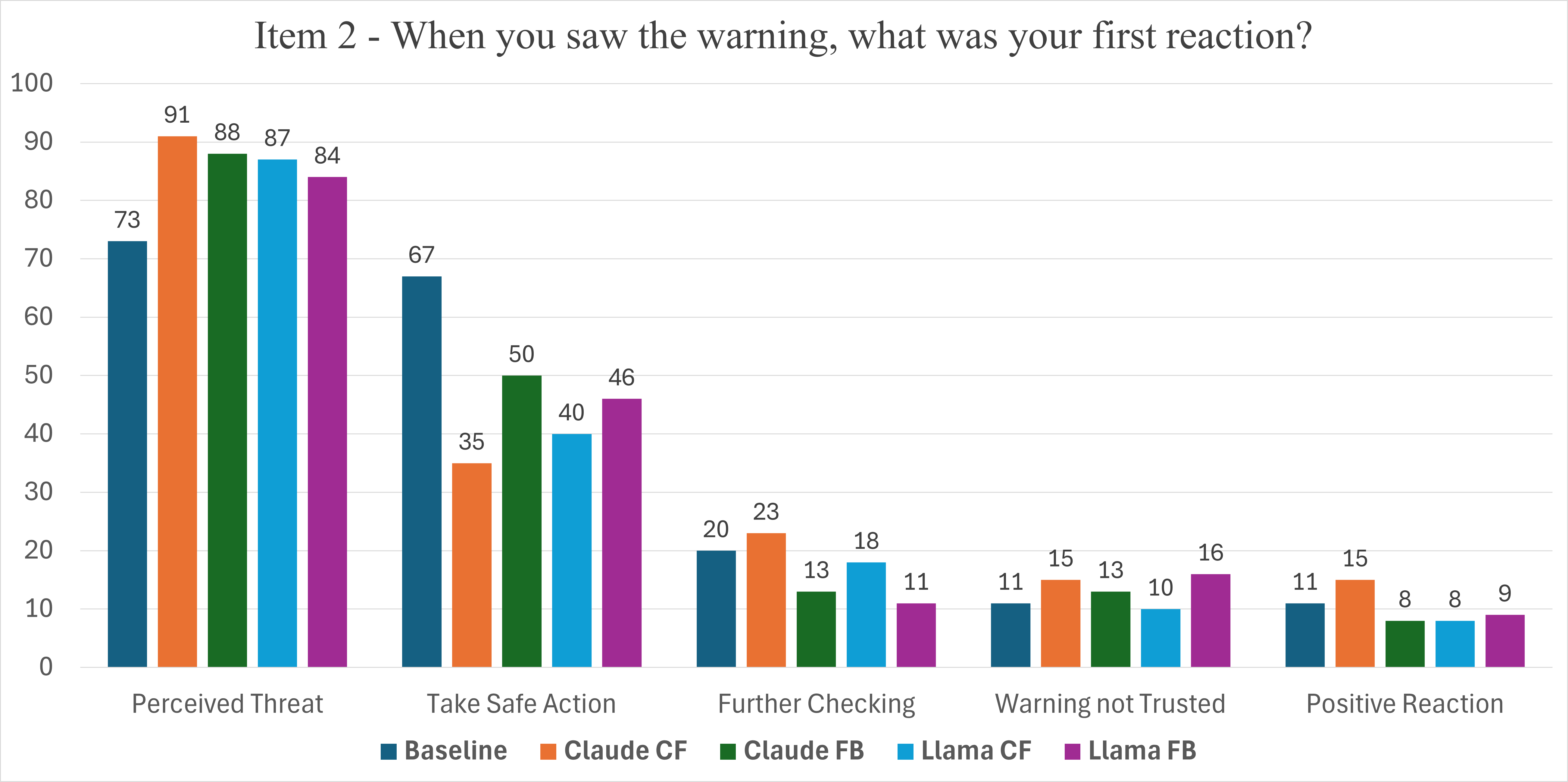}
    \caption{Distribution of first reaction themes (Item 2) across experimental conditions, including the manually crafted feature-based baseline from \cite{greco2025enhancing}, and the combination of LLMs (LLaMA, Claude) and explanation styles (feature-based, counterfactual). The total number of instances for each theme is shown atop each bar.}
    \label{fig:reaction_themes}
\end{figure*}

In particular, Figures \ref{fig:reaction_themes} and \ref{fig:warning_meaning} show the frequencies for Item 2 and Item 9 across our experimental conditions, while Table \ref{tab:confusing_words} reports the themes, counts, and participants' example responses emerged from the thematic analysis considering confusing words for Item 2.

\subsubsection{First reaction to warning (Item 2)}
In Item 2, participants answered the question: \enquote{When you saw the warning, what was your first reaction?}.
Five recurring themes were identified across the four experimental conditions: \textit{Perceived threat}, \textit{Take safe action}, \textit{Further checking}, \textit{Warning not trusted}, and \textit{Positive reaction to the warning}. These themes are reported in Figure~\ref{fig:reaction_themes}, which shows their frequency across the four experimental conditions defined by the LLM (LLaMA or Claude), the explanation style (feature-based or counterfactual), and the manual feature-based explanation from \cite{greco2025enhancing}.

\textbf{Perceived threat.} 
This theme captured participants’ immediate perception of danger, often accompanied by feelings of uncertainty or concern. For instance, one participant (Claude, feature-based) remarked: \textit{``It was strange, but I thought it had something to do with some particular links not being recognized by the computer''} (P290). Another (LLaMA, counterfactual) noted: \textit{``That the site I am potentially about to visit is malicious and may cause harm to my computer.''} (P393). Claude FB tends to elicit this perception more than the other conditions (91 times), whereas the baseline elicits this perception to a lesser extent (73 times).

\textbf{Take safe action.} 
Many users responded by choosing the safest path, typically avoiding interaction with the email or the suspicious link. A participant (Claude, counterfactual) stated: \textit{``To click on 'Back to safety'''}  (P9), while another (LLaMA, feature-based) wrote: \textit{``That the site I am about to visit is at a high risk of being a scam site and I am advised to be careful''} (P512). Here, it is evident that in the baseline condition, this action is elicited very often (67 times). In contrast, for LLM-generated conditions, this idea is much less considered by users (35 to 46 times).

\textbf{Further checking.} 
Some participants described their intention to verify the warning by inspecting the email or sender. For example, a participant (Claude, feature-based) commented: \textit{``Stop and think about what to do''} (P518), and another (LLaMA, counterfactual) shared: \textit{``I wanted to verify who the sender was and if the link made sense''} (P120). The feature-based conditions elicit this action less frequently (13 times Claude FB, 11 times Lama FB) than the other conditions (from 18 to 23 times).

\textbf{Warning not trusted.} 
A subset of responses expressed scepticism or confusion about the alert. One user (Claude, counterfactual) admitted: \textit{``I was a bit worried if this is not a real warning, but of course it is not.''}, while another (LLaMA, feature-based) stated: \textit{``I didn’t trust the warning, it seemed like a false alarm''} (P263). In this case, the results appear more balanced across all the experimental conditions.

\textbf{Positive reaction to the warning.} 
Some users welcomed the alert or found it informative. A participant (LLaMA, counterfactual) reflected: \textit{``It was helpful to get an alert like that, made me feel safer''} (P370). In this theme, the results are generally more balanced across all experimental conditions, except for the Claude FB, which elicited a positive reaction from participants (15 times).

\subsubsection{Confusing words (Item 6)}
With this item, we asked participants to indicate which words they found confusing or too technical in the warning dialogue with an open question. Overall, four themes emerged\footnote{Please note that 179 participants did not respond, as answering was optional.}, summarized in Table \ref{tab:confusing_words}.
The most common theme was \enquote{Warning is clear}, highlighting that participants did not find any word confusing or too technical. The second theme, \enquote{Unclear general word}, includes responses where participants reported confusion about general terms used in the warning text  (e.g., \enquote{deceptive}, \enquote{harmful}, or \enquote{suspicious}). 
The third theme, \enquote{Unclear technical term}, refers to participants who reported difficulties with technical terms included in the warning  (e.g., \enquote{IP address}, or \enquote{https}). The last theme is \enquote{Unclear UI element}, referring to participants who found the words and buttons of the warning dialogue misleading  (e.g., the function of the button \enquote{click here (not safe) to read} was sometimes misinterpreted by participants).

\definecolor{Gray}{gray}{0.95} 

\begin{table}[htbp]
\caption{Themes, counts, and example answers for confusing words item: \enquote{\textit{Which word(s) did you find confusing or too technical?}}}
\label{tab:confusing_words}

\rowcolors{2}{Gray}{white}
\begin{tabular}{p{2.8cm} r p{4.2cm}}
\toprule
\textbf{Theme} & \textbf{Count} & \textbf{Example answers} \\
\midrule
Warning is clear & 282 & \enquote{I didn't find any part too confusing or technical.} (P230) \\
Unclear general word & 109 & \enquote{The numbers provided in the texts, such as the email points} (P52); \enquote{The word \textit{Deceptive}} (P41) \\
Unclear technical term & 5 & \enquote{The word \textit{https}} (P230) \\
Unclear UI element & 13 & \enquote{To still have to click the link clearly specified \textit{not to be safe} at the end of the warning} (P11) \\
\bottomrule
\end{tabular}
\end{table}

\begin{figure*}[ht]
    \centering
    \includegraphics[width=0.75\textwidth]{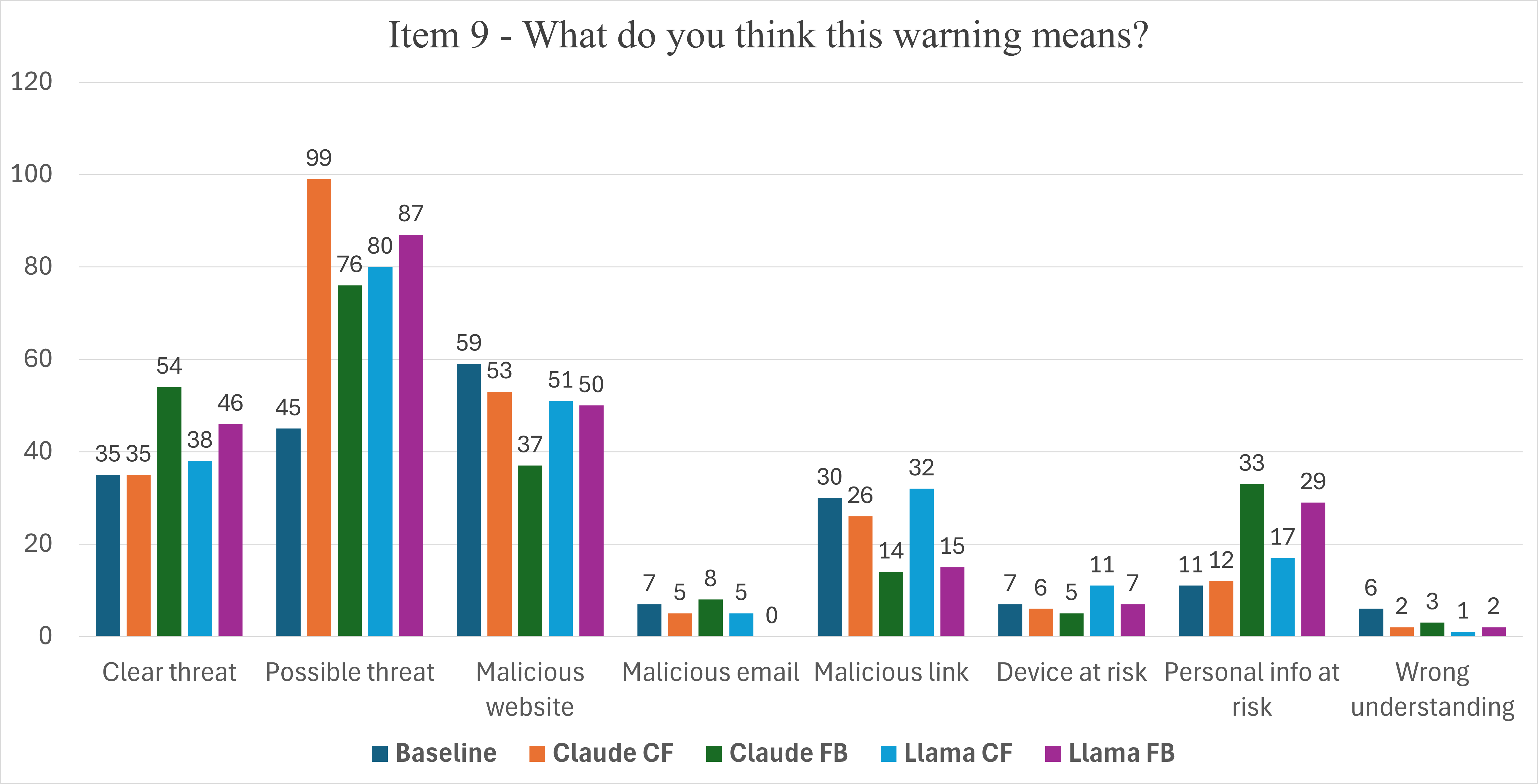}
    \caption{Distribution of warning meaning themes (Item 9) across experimental conditions, including the manually crafted feature-based baseline from \cite{greco2025enhancing}, and the combination of LLMs (LLaMA, Claude) and explanation styles (feature-based, counterfactual). The total number of instances for each theme is displayed at the top of each bar.}
    \label{fig:warning_meaning}
\end{figure*}

\subsubsection{Warning meaning (Item 9)}
In this item, participants responded to the question \enquote{What do you think this warning
means?}. Thematic analysis revealed eight distinct themes, presented in Fig. \ref{fig:warning_meaning}, along with their frequencies across the four conditions and the manually crafted feature-based explanation (baseline).

\textbf{Clear threat.}
Participants indicated the presence of a generic threat in the suspicious email or website, prompting them to avoid clicking on links or accessing the content (e.g., \enquote{It means that the website you're trying to enter is a scam and not real.
}, P77). Here, the feature-based conditions elicit this perception more frequently (54 times Claude FB and 46 times Lama FB) than the other conditions (from 35 to 38 times).

\textbf{Possible threat.}
Participants interpreted the warning as indicating that the content could be dangerous or lead to undesirable consequences, though not with certainty, prompting a need for further investigation (e.g., \enquote{It warns me against a possible phishing scam
}, P80). It is noticeable that the baseline, with a manually crafted feature-based explanation, instils less perceived threat (45 times) than LLM-generated explanations (from 76 to 99 times).

\textbf{Malicious website.}
Participants' interpretations of the warning varied, particularly concerning the specific source of danger. The threat was often associated with the website linked in the email. Additionally, some participants perceived the warning as indicating a particular threat (e.g., \enquote{ That the link takes you to a deceptive website and not to the one that was supposed to be.}, P85) while others saw it as more general (e.g., \enquote{ It's letting you know that it is fake and that you can potentially be hacked.}, P532). Participants had a similar impression across all conditions (from 50 to 59 times), except for Claude FB, which resulted in a lesser impression of a malicious website (37 times).

\textbf{Malicious email.}
In some cases, participants interpreted the warning as indicating that the threat resided in the content of the email itself or a scammer sender. Interpretations within this theme varied in terms of perceived threat certainty, ranging from a definitive assessment of maliciousness (e.g., \enquote{It means that the email has been identified as being sent from a scammer.}, P244) to more tentative assessments of potential risk (e.g., \enquote{Warns users to be careful of suspicious links/emails that can take their data}, P214). This theme was mentioned infrequently across conditions.

\textbf{Malicious link.}
In some instances, participants understood the warning as indicating that the threat was embedded in the link. The perceived level of risk varied, ranging from a clear and immediate danger (e.g., \enquote{That the link I clicked leads to a site that doesn't belong to Facebook and they will steal the information I type.}, P361) to a more cautious interpretation of potential harm (e.g., \enquote{Warns users to be careful of suspicious links/emails that can take their data}, P215). This theme was more frequent in counterfactual (26 times for Claude CF and 32 times for Llama CF) and baseline explanations than in feature-based (14 times for Claude FB and 15 times for Llama FB) explanations.

\textbf{Device at risk.}
The warning was interpreted as an indication that the user's device could be at risk, for example, due to potential malware contained within the email content or linked website (e.g., \enquote{It means the webpage or link you are about to enter may harm your computer or breach data}, P513). This theme was generally less frequent than others and quite balanced across all conditions. 

\textbf{Personal information at risk.}
Some participants interpreted the warning as suggesting that their data, such as login credentials, were at risk of being stolen by cybercriminals (e.g., \enquote{This warning means the website you're attempting to visit is likely a fake site that's designed to steal login details.}, P157). This theme was more frequent for feature-based explanations (32 times in the case of Claude FB and 29 times for Llama FB) than for the baseline and counterfactual ones (12 times in the case of Claude CF and 17 times for Llama CF).

\textbf{Wrong understanding.}
The final theme encompasses all instances where participants misinterpreted the warning (e.g., \enquote{It means the website has a bug}, P131; \enquote{That I can not continue with whatever I was doing then}, P326). Overall, this was the most infrequent theme, where the manually crafted explanations (baseline) achieved the highest counts.

\subsubsection{Impact of Comprehension on Decision Making}
Beyond identifying isolated themes, we cross-referenced these qualitative insights with quantitative behavioral metrics to validate the causal link between explanation content and user action. Notably, users in the Claude condition who avoided the phishing link frequently cited specific details from the generated text (e.g., "The warning mentioned a mismatch in the URL") in their open-ended responses. Conversely, users who clicked often dismissed the warning as a "false alarm" rather than misunderstanding the text. This triangulation confirms that the \textit{content} of the explanation, and not just the presence of a warning, influenced the decision-making process, validating the intervention's effectiveness beyond the generic effects of a warning, such as those without explanations.

\subsection{Linking Comprehension to Behavior}
To address the causality between explanation content and user action, we cross-referenced qualitative feedback with quantitative behavioral metrics. Notably, users in the Claude condition who avoided the phishing link frequently cited specific details from the explanation (e.g., "The warning mentioned a mismatch in the URL") in their open-ended responses. Conversely, users who clicked often dismissed the warning as a "false alarm" rather than misunderstanding the text. This suggests that the \textit{content} of the explanation directly influenced the decision-making process, validating the intervention's effectiveness beyond generic warning effects.

\section{Threats to validity} \label{sec:threats_validity}

In this section, we analyse some potential issues that might threaten the study's validity. 

\subsection{Internal Validity}
Internal validity can be compromised by hidden factors that may threaten the conclusions drawn. 

\textbf{Confounding Variables.}
Latent factors, such as prior knowledge and experience with phishing, may have influenced behavior and responses to warnings. We employed a between-subjects design to mitigate this threat, and participants were randomly assigned to one of the conditions. 

\textbf{Deception and priming.}
At the beginning of the experiment, participants were informed that the study aimed to evaluate a new email client interface, thereby concealing the true objective of the study, which was revealed at the end to avoid biasing participants towards unnatural behavior. Even if this procedure was ethically approved to reduce demand characteristics, such framing may have influenced attentiveness and the perceived seriousness of the task.

\subsection{External Validity}

External validity regards the extent to which the results can be generalized to broader contexts, populations, and real-world settings.

\textbf{Ecological validity.}
The email client we created mimics real and common features of web clients; however, users were aware that they were participating in a study, thus in a safe environment. This reduced-risk setting can lead to unnatural behaviours that users exhibit when facing phishing attacks. This risk was mitigated by asking users to behave as naturally as possible. 

\textbf{Participant recruiting.}
To make the study population as representative as possible of a generic population affected by phishing attacks, we used Prolific by setting only gender and English proficiency as inclusion criteria. This guaranteed that participants were recruited worldwide and of different ages. However, considering the typical Prolific users, our sample may not include older adults, cybersecurity professionals, or individuals with low digital literacy. 

\textbf{Device and context effects.}
Participants attended the study remotely on their devices in uncontrolled environments. Factors such as device, screen size, multitasking, or ambient noise may have affected attention and decision-making.

\subsection{Construct Validity}

Construct validity examines whether the study accurately measures its intended concepts, namely, the effectiveness and perception of LLM-generated phishing explanations.

\textbf{Behavioral measures.}
We adopted the click-through rate (CTR), the most widely adopted metric to measure phishing vulnerability. However, clicks may occur in a controlled and safe setting for various reasons, including curiosity or misunderstanding. Thus, this behaviour may not always reflect a failure to detect phishing intent.

\subsection{Statistical Validity}

Statistical validity refers to the appropriateness of the data analysis methods and the robustness of the inferences drawn.

\textbf{Multiple comparisons.}
In our study, various statistical tests, including Chi-square analyses, Kruskal-Wallis tests, and logistic regressions, were conducted. However, while the Bonferroni correction was applied where applicable, the risk of Type I errors was reduced, but it marginally remained.

\section{Lessons learned} \label{sec:lessons_learned}
Triangulating qualitative and quantitative findings, we distilled lessons learned and provided direct guidance on designing and implementing future decisions. The four research questions are correlated with these lessons, thus determining the practical implications of our studies. They explain how the interaction between various design options, methods of explanation, and user differences affects the effectiveness and the perceived utility of AI-produced phishing warning messages.

\subsection*{Lesson 1 (RQ1): The Economic Case for Automated Explanations}

Our results provide empirical evidence that LLM-generated explanations—specifically those from Claude 3.5 Sonnet—achieve protection levels that are \textit{statistically indistinguishable} from expert-crafted messages (p $>$ 0.05). Crucially, the observed click-through rate for Claude (9.39\%) was numerically lower than the manual baseline (9.94\%), indicating that automation introduces no performance degradation.

This finding carries substantial \textit{practical significance}. Manually crafting context-aware explanations for every potential phishing vector is unscalable and cost-prohibitive. Conversely, our results demonstrate that LLMs can satisfy a requirement of \textit{operational non-inferiority} purely through automated prompting. This shifts the paradigm from asking "Are LLMs significantly better than experts?" to recognizing that "LLMs are sufficiently capable to democratize high-quality security explanations at scale without compromising safety." These findings align with recent work suggesting that properly fine-tuned LLM outputs can match human efforts in causal reasoning \citep{Liao2020CHI}, extending this validity to the critical domain of phishing defense.

\textbf{Takeaway.} 
Security practitioners can leverage high-quality LLMs to automate warning generation. Since the automated output proves to be statistically comparable to the manual "gold standard," the focus should shift from content creation to prompt engineering and validation, enabling scalable, real-time defenses against evolving threats.

\subsection*{Lesson 2 (RQ2): Feature-based explanations may enhance the accuracy in identifying actual phishing emails.}

In investigating feature-based explanation styles, the present study reveals that in the case of true positive emails, feature-based resulted in a lower CTR (11.03\%) with respect to counterfactual ones (12.04\%). This difference was more pronounced in the case of Claude-generated explanations (Claude FB: 7.48\% vs. Claude CF: 10.27\%), although this should be interpreted as an observed trend rather than a statistically significant effect, given the p-value exceeding conventional thresholds. These insights are consistent with the recent study by \citet{zhao2024explainability}, which shows that slight differences in language model tuning can trigger strong behavioral outcomes. The analysis of Item 9 further illustrates that explanations issued by Claude more frequently prompted interpretations invoking data theft or credential loss, thus suggesting a heightened salience of risk.

\textbf{Takeaway.} 
Feature-based explanations may help reduce clicks on phishing attempts, although further research is needed to investigate this trend more deeply; in particular, LLMs such as Claude appear to convey a greater sense of threat.

\subsection{Lesson 3 (RQ2): Counterfactuals may reduce false alarm overrides}

Even if counterfactuals showed no overall CTR advantage, it emerged that they might reduce false positive overrides (10.00\% vs. 14.06\% for feature-based). Additionally, qualitative responses (e.g., “The email would have been safe if…” [P393]) suggest that counterfactuals help users distinguish true threats from benign anomalies. Thus, counterfactuals help users reason about \emph{why} the content may or may not be dangerous.

\textbf{Takeaway.} 
Adopt counterfactual explanations when the risk of false positives is high, such as in sensitive domains (e.g., internal communication or finance), where false alarms can disrupt workflows.

\subsection*{Lesson 4 (RQ2, RQ3): Explanation style shapes the user's mental model of phishing risk}

The analysis of Item 9 revealed that feature-based and counterfactual explanations elicit different sources of threats in users. In the case of feature-based explanations, users tend to identify the outcome of the threat, for example, stolen credentials or leaked data. On the contrary, in the case of counterfactual explanations, users often pinpoint the mechanism of deception, particularly with deceptive links and their mismatch with related labels. This finding aligns with findings in the literature \cite{djatsa2020threat, greco2025enhancing}, which reveal that the explanations of the threat influence users' beliefs about where the risk resides.

\textbf{Takeaway.} Counterfactual explanations should be presented to highlight manipulation tactics, while feature-based explanations should be visualized to reinforce the severity of potential consequences.

\subsection*{Lesson 5: The Readability-Security Paradox}
 
Our findings reveal a counterintuitive divergence between perceived understandability and behavioral effectiveness, a phenomenon we term the \textit{Readability-Security Paradox}. Llama 3.3 generated the most "understandable" warnings (statistically significant vs. baseline, $p=0.002$) yet resulted in higher click-through rates. In contrast, Claude 3.5 produced more complex explanations—linking features to consequences, such as "identity theft"—and achieved better protection despite lower readability scores. This suggests that \textit{clarity alone is insufficient} for security warnings. Excessive simplification or a neutral tone (characteristic of Llama's outputs) might inadvertently lower the user's risk perception, leading to a "compliance" failure despite "comprehension" success. 
Effective security warnings may require a degree of "cognitive friction" or semantic severity that purely readable text fails to convey. Future LLM prompting strategies should therefore prioritize "risk salience" over simple "readability", tuning models to balance clarity with \textit{fear appeals} or consequence-based framing.

\textbf{Takeaway.} 
Do not optimize solely for readability metrics. LLM prompts should be engineered to balance linguistic clarity with \textit{risk salience}. Instruct models to explicitly articulate severe consequences (e.g., data theft) rather than maintaining a neutral tone, thereby introducing the necessary cognitive friction to drive compliant behavior.

\subsection*{Lesson 6 (RQ4): Habituation and high online exposure both increase risk}

The study results suggest that familiarity with warnings and the extended use of the internet daily are both highly significant predictors of clicking on phishing emails. In particular, the odds of clicking were more than doubled by a one-point increase in familiarity (OR = 2.20, p = .001), which indicates that habituation is a direct detriment to the effectiveness of warnings. Similarly, each extra hour spent online per day increased click probability by about 25\% (OR = 1.25, p = .024), implying that individuals exposed to a lot of online content have a high chance of adopting automatic or shortcut browsing behaviours.  
These findings are consistent with prior studies on habituation in the warning design literature (\citealp{amran2018habituation}, \citealp{akhawe2013alice}, \citealp{egelman2008warned}), which have shown that familiarity may reduce people's sensitivity to a familiar warning signal.

\textbf{Takeaway.} 
Warning designs should disrupt habitual browsing by introducing visual or semantic variability, especially for users with high exposure to online content. For example, designers can rotate, refresh, or personalise warning designs to sustain user attention and reduce the risks posed by the habituation effect.

\subsection*{Lesson 7 (RQ4): Mind the (cognitive) gap -- workload slows risky clicks}

The data from the NASA-TLX questionnaire revealed a significant negative correlation between the perceived workload and total click-through rate ($\beta = -0.029$, $p=0.002$). In particular, the increased mental effort was associated with a decline in the probability of impulsive clicks. This observation was supported via logistic regression models, which showed that the odds of clicking reduced by 34\% (OR=0.66, $p=0.012$) with each increment of workload. Further, the impact was similar across the explanations and data subsets styles, such as true-positive (OR=0.65, $p=0.012$) and false-positive conditions (e.g., Claude aggregate FP; OR=0.53, $p=0.004$).

The findings are congruent with past results by 
\citep{sheahan2024designing} who state that constructive friction promotes reflection before doing something. Our finding is also in line with 
\citep{Kahneman2011Thinking}, which illustrates how shifting from fast (impulsive) to slow (deliberative) thinking improves judgment in risk-laden contexts.

\textbf{Takeaway.} Introducing mild cognitive friction, such as a prompt or delay, might help prevent impulsive decisions on phishing links.

\subsection*{Lesson 8 (RQ4): Gendered clicks - demographics matter}

Logistic regressions revealed that gender had a significant and systematic effect on click behavior. The likelihood of female participants clicking on phishing emails was significantly lower, and the odds ratios averaged almost zero in true-positive and false-positive phishing email conditions. On the contrary, male participants with higher values of Need for Cognition (NFC) were over twice as likely to click in the false-positive conditions with counterfactual warnings (OR = 2.16, p = .012). The gender-by-NFC interaction suggests that certain cognitive characteristics, in combination with gender, may increase vulnerability, possibly due to an over-trust in the reasoning ability when exposed to ambiguous cues by the user. These results support the findings obtained by 
\cite{greitzer2021experimental}, who also observed that male participants were more susceptible to phishing attacks in simulated attacks; they explained the trend with increased overconfidence or lowered caution.

\textbf{Takeaway.} Security messaging may benefit from adaptive targeting that accounts for demographic factors like gender.

\subsection*{Lesson 9: Safety Guardrails and the Risk of Unconstrained Generation}

The selection phase of the LLM models revealed a critical insight: \textit{unconstrained LLMs}, i.e., models generating text without external verification layers, pose significant security risks. In our preliminary tests, models such as Gemini 1.5 Flash and Llama 3.1 8B hallucinated when prompted directly, generating nonexistent URL features or providing vague safety assurances for malicious links. 
This result indicates that purely generative approaches are insufficient for critical defense tasks. To ensure safety and mitigate this risk, our study employed a \textit{Human-in-the-Loop} validation approach. Similarly, the real-world deployment of the final system must consider a "hybrid architecture" where generative outputs are constrained by deterministic checks (e.g., code that verifies whether the explained URL actually matches the email header) before being presented to the user.

\textbf{Takeaway.} 
Generative AI for warning dialogs requires a "trust but verify" approach. Outputs generated by LLMs must be rigorously validated to prevent hallucinations, ensuring that false assurances do not paradoxically lower user protection.

\section{Conclusions} \label{sec:conclusions}
This study provides empirical evidence that LLMs can automatically generate phishing warning explanations that are as effective as, and in certain cases, more effective than those crafted by humans. Notably, warnings generated by Claude 3.5 Sonnet markedly reduced click-through rates on phishing links compared with Llama 3.3 70B and human-composed messages, especially when a feature-based explanation style was used. Counterfactual explanations initially showed potential in reducing false positives; however, feature-based explanations remained more reliable for identifying genuine threats.

The viability of LLM-generated warnings was supported with user-perception measures: participants found them clear, trustworthy, and actionable, especially those generated by Claude. The study also found that mental workload, gender, and familiarity with warning dialogues significantly affected the warning efficacy. These results emphasize the importance of individualization and flexibility in security interfaces.

Overall, employing LLMs in phishing defense systems can improve scalability, responsiveness, and user interaction without reducing effectiveness. Future work should examine real-time deployment, user-personalisation strategies, and broader applications of LLM-generated explanations in cybersecurity interfaces.


\section*{Declaration of generative AI}
During the writing of this paper, the author(s) used \textit{Grammarly Pro} to fix grammatical errors and improve text quality. After using this tool/service, the author(s) reviewed and edited the content as needed and take(s) full responsibility for the content of the published article.

\section*{Data Availability Statement}

All materials supporting the findings of this study are publicly available in the following repository:

\begin{center}
\url{https://github.com/IVU-Laboratory/llm_warnings_explantions}
\end{center}

The repository includes: (i) the complete source code of the experimental web platform used to conduct the user study; (ii) detailed documentation of the explanation-generation process, including all feature-based and counterfactual explanations evaluated prior to the user study; (iii) screenshots of all email stimuli presented to participants during the experiment; and (iv) raw, anonymized interaction data together with processed datasets and scripts used for the statistical analyses reported in this paper.

All user data have been fully anonymized, and no personally identifiable information is included. The shared materials enable full transparency and support the reproducibility of the experimental design, data analysis, and reported results.

\section*{Acknowledgments}
This work has been supported by the Italian Ministry of University and Research (MUR) and by the European Union-NextGenerationEU, under grant PRIN 2022 PNRR "DAMOCLES: Detection And Mitigation Of Cyber attacks that exploit human vuLnerabilitiES" (Grant P2022FXP5B) CUP: H53D23008140001. 
    
This work is partially supported by the co-funding of the European Union - Next Generation EU: NRRP Initiative, Mission 4, Component 2, Investment 1.3 - Partnerships extended to universities, research centres, companies and research D.D. MUR n. 341 del 5.03.2022 – Next Generation EU (PE0000014 – "Security and Rights In the CyberSpace – SERICS" - CUP: H93C22000620001). 
    
The research of Francesco Greco is funded by a PhD fellowship within the framework of the Italian "D.M. n. 352, April 9, 2022"- under the National Recovery and Resilience Plan, Mission 4, Component 2, Investment 3.3 - PhD Project "Investigating XAI techniques to help user defend from phishing attacks", co-supported by "Auriga S.p.A." (CUP H91I22000410007).

\bibliographystyle{abbrvnat} 
\bibliography{sample-base}

@article{Khonji2013Phishing,
  title={Phishing detection: a literature survey},
  author={Khonji, Mahmoud and Iraqi, Youssef and Jones, Andrew},
  journal={IEEE Communications Surveys \& Tutorials},
  volume={15},
  number={4},
  pages={2091--2121},
  year={2013},
  publisher={IEEE},
  doi={10.1109/SURV.2013.032213.00009},
  url={https://doi.org/10.1109/SURV.2013.032213.00009}
}

@inproceedings{Prakash2010PhishNet,
  author={Prakash, Pawan and Kumar, Manish and Kompella, Ramana Rao and Gupta, Minaxi},
  booktitle={2010 Proceedings IEEE INFOCOM}, 
  title={PhishNet: Predictive Blacklisting to Detect Phishing Attacks}, 
  year={2010},
  volume={},
  number={},
  pages={1-5},
  doi={10.1109/INFCOM.2010.5462216},
  url={https://doi.org/10.1109/INFCOM.2010.5462216}
}

@article{Guidotti2019Counterfactual,
  author={Guidotti, Riccardo and Monreale, Anna and Giannotti, Fosca and Pedreschi, Dino and Ruggieri, Salvatore and Turini, Franco},
  journal={IEEE Intelligent Systems}, 
  title={Factual and Counterfactual Explanations for Black Box Decision Making}, 
  year={2019},
  volume={34},
  number={6},
  pages={14-23},
  doi={10.1109/MIS.2019.2957223}
}

@misc{Xu2025LLMsCybersecurity,
      title={Large Language Models for Cyber Security: A Systematic Literature Review}, 
      author={Hanxiang Xu and Shenao Wang and Ningke Li and Kailong Wang and Yanjie Zhao and Kai Chen and Ting Yu and Yang Liu and Haoyu Wang},
      year={2025},
      eprint={2405.04760},
      archivePrefix={arXiv},
      primaryClass={cs.CR},
      url={https://arxiv.org/abs/2405.04760}, 
      doi={10.48550/arXiv.2405.04760}
}

@Article{Charmet2022XAI4Cybersecurity,
author={Charmet, Fabien
and Tanuwidjaja, Harry Chandra
and Ayoubi, Solayman
and Gimenez, Pierre-Fran{\c{c}}ois
and Han, Yufei
and Jmila, Houda
and Blanc, Gregory
and Takahashi, Takeshi
and Zhang, Zonghua},
title={Explainable artificial intelligence for cybersecurity: a literature survey},
journal={Annals of Telecommunications},
year={2022},
month={Dec},
day={01},
volume={77},
number={11},
pages={789-812},
issn={1958-9395},
doi={10.1007/s12243-022-00926-7},
url={https://doi.org/10.1007/s12243-022-00926-7}
}

@ARTICLE{Rjoub2023XAI4Cybersecurity,
  author={Rjoub, Gaith and Bentahar, Jamal and Abdel Wahab, Omar and Mizouni, Rabeb and Song, Alyssa and Cohen, Robin and Otrok, Hadi and Mourad, Azzam},
  journal={IEEE Transactions on Network and Service Management}, 
  title={A Survey on Explainable Artificial Intelligence for Cybersecurity}, 
  year={2023},
  volume={20},
  number={4},
  pages={5115-5140},
  doi={10.1109/TNSM.2023.3282740}
 }

@article{greco2025enhancing,
title = {Enhancing Phishing Defenses: The Impact of Timing and Explanations in Warnings for Email Clients},
journal = {Computer Standards \& Interfaces},
volume = {93},
pages = {103982},
year = {2025},
issn = {0920-5489},
doi = {https://doi.org/10.1016/j.csi.2025.103982},
url = {https://www.sciencedirect.com/science/article/pii/S092054892500011X},
author = {Francesco Greco and Giuseppe Desolda and Paolo Buono and Antonio Piccinno}
}

@article{SARKER2024XAI4CS,
title = {Explainable AI for cybersecurity automation, intelligence and trustworthiness in digital twin: Methods, taxonomy, challenges and prospects},
journal = {ICT Express},
volume = {10},
number = {4},
pages = {935-958},
year = {2024},
issn = {2405-9595},
doi = {https://doi.org/10.1016/j.icte.2024.05.007},
url = {https://www.sciencedirect.com/science/article/pii/S2405959524000572},
author = {Iqbal H. Sarker and Helge Janicke and Ahmad Mohsin and Asif Gill and Leandros Maglaras}
}

@article{desolda2023explanations,
title = {Explanations in warning dialogs to help users defend against phishing attacks},
journal = {International Journal of Human-Computer Studies},
volume = {176},
pages = {103056},
year = {2023},
issn = {1071-5819},
doi = {https://doi.org/10.1016/j.ijhcs.2023.103056},
url = {https://www.sciencedirect.com/science/article/pii/S1071581923000654},
author = {Giuseppe Desolda and Joseph Aneke and Carmelo Ardito and Rosa Lanzilotti and Maria Francesca Costabile}
}

@article{desolda2025apollo,
    author = "Giuseppe Desolda and Francesco Greco and Luca Vigano",
    title = "APOLLO: A GPT-based tool to detect phishing emails and generate explanations that warn users",
    journal={Proceedings of the ACM on Human-Computer Interaction},
    volume = "9",
    issue = "4",
    article = "EICS003",
    date = "June 2025",
    pages = "33",
    doi = "10.1145/3733049",
    url = "https://doi.org/10.1145/3733049",
    year = "2025",
    publisher = {ACM New York, NY}
}

@article{macefield2007usability,
  title={Usability studies and the Hawthorne Effect},
  author={Macefield, Ritch},
  journal={Journal of usability studies},
  volume={2},
  number={3},
  pages={145--154},
  year={2007},
  publisher={Usability Professionals' Association Bloomingdale, IL}
}

@inproceedings{nicholson2017can,
  title={Can we fight social engineering attacks by social means? Assessing social salience as a means to improve phish detection},
  author={Nicholson, James and Coventry, Lynne and Briggs, Pam},
  booktitle={Thirteenth Symposium on Usable Privacy and Security (SOUPS 2017)},
  pages={285--298},
  year={2017}
}

@inproceedings{chitare2023may,
  title={“It may take ages”: Understanding Human-Centred Lateral Phishing Attack Detection in Organisations},
  author={Chitare, Neeranjan and Coventry, Lynne and Nicholson, James},
  booktitle={Proceedings of the 2023 European Symposium on Usable Security},
  pages={344--355},
  year={2023}
}

@article{chitare2025exploring,
  title={Exploring transparent communication for organisational cyber-resilience to sophisticated phishing attacks},
  author={Chitare, Neeranjan and Coventry, Lynne and Nicholson, James},
  journal={Information \& Computer Security},
  year={2025},
  publisher={Emerald Publishing Limited}
}

@misc{Murdoch2017,
  author       = {Murdoch, Steven J. and Sasse, Angela},
  title        = {Should you phish your own employees?},
  howpublished = {Bentham's Gaze},
  year         = {2017},
  month        = aug,
  day          = {22},
  url          = {https://www.benthamsgaze.org/2017/08/22/should-you-phish-your-own-employees/},
  note         = {Accessed: 2025-10-15}
}

@Article{Basit2021Survey,
author={Basit, Abdul
and Zafar, Maham
and Liu, Xuan
and Javed, Abdul Rehman
and Jalil, Zunera
and Kifayat, Kashif},
title={A comprehensive survey of AI-enabled phishing attacks detection techniques},
journal={Telecommunication Systems},
year={2021},
month={Jan},
day={01},
volume={76},
number={1},
pages={139-154},
issn={1572-9451},
doi={10.1007/s11235-020-00733-2},
url={https://doi.org/10.1007/s11235-020-00733-2}
}

@article{McLaughlin1969SMOG,
 ISSN = {00224103},
 author = {G. Harry Mc Laughlin},
 journal = {Journal of Reading},
 number = {8},
 pages = {639--646},
 publisher = {[Wiley, International Reading Association]},
 title = {SMOG Grading-a New Readability Formula},
 urldate = {2025-05-19},
 url = {http://www.jstor.org/stable/40011226},
 volume = {12},
 year = {1969}
}

@book{kincaid1975derivation,
  title={Derivation of New Readability Formulas: (automated Readability Index, Fog Count and Flesch Reading Ease Formula) for Navy Enlisted Personnel},
  author={Kincaid, J.P.},
  series={Research Branch report},
  url={https://books.google.it/books?id=4tjroQEACAAJ},
  year={1975},
  publisher={Chief of Naval Technical Training, Naval Air Station Memphis}
}

@article{Samed2025XAI4CSIntrusionDetectionSystems,
title = {Explainable artificial intelligence models in intrusion detection systems},
journal = {Engineering Applications of Artificial Intelligence},
volume = {144},
pages = {110145},
year = {2025},
issn = {0952-1976},
doi = {https://doi.org/10.1016/j.engappai.2025.110145},
url = {https://www.sciencedirect.com/science/article/pii/S0952197625001459},
author = {Samed AL and Seref Sagiroglu}
}

@inproceedings{Cau25MultiExp,
author = {Cau, Federico Maria and Spano, Lucio Davide},
title = {The Influence of Curiosity Traits and On-Demand Explanations in AI-Assisted Decision-Making},
year = {2025},
isbn = {9798400713064},
publisher = {ACM},
address = {New York, NY, USA},
url = {https://doi.org/10.1145/3708359.3712165},
doi = {10.1145/3708359.3712165},
booktitle = {Proceedings of the 30th International Conference on Intelligent User Interfaces},
pages = {1440–1457},
numpages = {18},
location = {
},
series = {IUI '25}
}

@article{Fok24FeatureBased,
author = {Fok, Raymond and Weld, Daniel S.},
title = {In search of verifiability: Explanations rarely enable complementary performance in AI‐advised decision making},
year = {2024},
issue_date = {Fall 2024},
publisher = {John Wiley \& Sons, Inc.},
address = {USA},
volume = {45},
number = {3},
issn = {0738-4602},
url = {https://doi.org/10.1002/aaai.12182},
doi = {10.1002/aaai.12182},
journal = {AI Mag.},
month = sep,
pages = {317–332},
numpages = {16}
}

@article{Koh24AlgorithmicRecourse,
author = {Koh, Seunghun and Kim, Byung Hyung and Jo, Sungho},
title = {Understanding the User Perception and Experience of Interactive Algorithmic Recourse Customization},
year = {2024},
issue_date = {June 2024},
publisher = {ACM},
address = {New York, NY, USA},
volume = {31},
number = {3},
issn = {1073-0516},
url = {https://doi.org/10.1145/3674503},
doi = {10.1145/3674503},
journal = {ACM Transaction Computer-Human Interaction},
month = aug,
articleno = {43},
numpages = {25}
}

@article{Verma24AlgorithmicRecourse,
author = {Verma, Sahil and Boonsanong, Varich and Hoang, Minh and Hines, Keegan and Dickerson, John and Shah, Chirag},
title = {Counterfactual Explanations and Algorithmic Recourses for Machine Learning: A Review},
year = {2024},
issue_date = {December 2024},
publisher = {ACM},
address = {New York, NY, USA},
volume = {56},
number = {12},
issn = {0360-0300},
url = {https://doi.org/10.1145/3677119},
doi = {10.1145/3677119},
journal = {ACM Comput. Surv.},
month = oct,
articleno = {312},
numpages = {42}
}

@inproceedings{Upadhyay25AlgorithmicRecourse,
author = {Upadhyay, Sohini and Lakkaraju, Himabindu and Gajos, Krzysztof Z.},
title = {Counterfactual Explanations May Not Be the Best Algorithmic Recourse Approach},
year = {2025},
isbn = {9798400713064},
publisher = {ACM},
address = {New York, NY, USA},
url = {https://doi.org/10.1145/3708359.3712095},
doi = {10.1145/3708359.3712095},
booktitle = {Proceedings of the 30th International Conference on Intelligent User Interfaces},
pages = {446–462},
numpages = {17},
location = {
},
series = {IUI '25}
}

@article{Lee23Counterfactual,
author = {Lee, Min Hun and Chew, Chong Jun},
title = {Understanding the Effect of Counterfactual Explanations on Trust and Reliance on AI for Human-AI Collaborative Clinical Decision Making},
year = {2023},
issue_date = {October 2023},
publisher = {ACM},
address = {New York, NY, USA},
volume = {7},
number = {CSCW2},
url = {https://doi.org/10.1145/3610218},
doi = {10.1145/3610218},
journal = {Proc. ACM Hum.-Comput. Interact.},
month = oct,
articleno = {369},
numpages = {22}
}

@article{Gentile25Counterfactual,
author = {Gentile, Davide and Donmez, Birsen and Jamieson, Greg A.},
title = {Human performance effects of combining counterfactual explanations with normative and contrastive explanations in supervised machine learning for automated decision assistance},
year = {2025},
issue_date = {Feb 2025},
publisher = {Academic Press, Inc.},
address = {USA},
volume = {196},
number = {C},
issn = {1071-5819},
url = {https://doi.org/10.1016/j.ijhcs.2024.103434},
doi = {10.1016/j.ijhcs.2024.103434},
journal={International Journal of Human-Computer Studies},
month = feb,
numpages = {11}
}

@inproceedings{VanNostrand24CounterfactualActionableRecourse,
author = {VanNostrand, Peter M. and Hofmann, Dennis M. and Ma, Lei and Rundensteiner, Elke A.},
title = {Actionable Recourse for Automated Decisions: Examining the Effects of Counterfactual Explanation Type and Presentation on Lay User Understanding},
year = {2024},
isbn = {9798400704505},
publisher = {ACM},
address = {New York, NY, USA},
url = {https://doi.org/10.1145/3630106.3658997},
doi = {10.1145/3630106.3658997},
booktitle = {Proceedings of the 2024 ACM Conference on Fairness, Accountability, and Transparency},
pages = {1682–1700},
numpages = {19},
location = {Rio de Janeiro, Brazil},
series = {FAccT '24}
}

@inproceedings{Lai2023SurveyHumanAIDecisionMaking,
author = {Lai, Vivian and Chen, Chacha and Smith-Renner, Alison and Liao, Q. Vera and Tan, Chenhao},
title = {Towards a Science of Human-AI Decision Making: An Overview of Design Space in Empirical Human-Subject Studies},
year = {2023},
isbn = {9798400701924},
publisher = {ACM},
address = {New York, NY, USA},
url = {https://doi.org/10.1145/3593013.3594087},
doi = {10.1145/3593013.3594087},
booktitle = {Proceedings of the 2023 ACM Conference on Fairness, Accountability, and Transparency},
pages = {1369–1385},
numpages = {17},
location = {<conf-loc>, <city>Chicago</city>, <state>IL</state>, <country>USA</country>, </conf-loc>},
series = {FAccT '23}
}

@inproceedings{Zhang2020ConfidenceExplanationsAccuracyTrust,
author = {Zhang, Yunfeng and Liao, Q. Vera and Bellamy, Rachel K. E.},
title = {Effect of confidence and explanation on accuracy and trust calibration in AI-assisted decision making},
year = {2020},
isbn = {9781450369367},
publisher = {ACM},
address = {New York, NY, USA},
url = {https://doi.org/10.1145/3351095.3372852},
doi = {10.1145/3351095.3372852},
booktitle = {Proceedings of the 2020 Conference on Fairness, Accountability, and Transparency},
pages = {295–305},
numpages = {11},
location = {Barcelona, Spain},
series = {FAT* '20}
}

@inproceedings{Wang2021ExampleBasedFeatureBasedAndOthers,
author = {Wang, Xinru and Yin, Ming},
title = {Are Explanations Helpful? A Comparative Study of the Effects of Explanations in AI-Assisted Decision-Making},
year = {2021},
isbn = {9781450380171},
publisher = {ACM},
address = {New York, NY, USA},
url = {https://doi.org/10.1145/3397481.3450650},
doi = {10.1145/3397481.3450650},
booktitle = {26th International Conference on Intelligent User Interfaces},
pages = {318–328},
numpages = {11},
location = {College Station, TX, USA},
series = {IUI '21}
}

@article{Chen2023RelianceExampleBasedFeatureBased,
author = {Chen, Valerie and Liao, Q. Vera and Wortman Vaughan, Jennifer and Bansal, Gagan},
title = {Understanding the Role of Human Intuition on Reliance in Human-AI Decision-Making with Explanations},
year = {2023},
issue_date = {October 2023},
publisher = {ACM},
address = {New York, NY, USA},
volume = {7},
number = {CSCW2},
url = {https://doi.org/10.1145/3610219},
doi = {10.1145/3610219},
journal = {Proc. ACM Hum.-Comput. Interact.},
month = {oct},
articleno = {370},
numpages = {32}
}

@inproceedings{Ma2023CorrectnessLikelihoodAIUsers,
author = {Ma, Shuai and Lei, Ying and Wang, Xinru and Zheng, Chengbo and Shi, Chuhan and Yin, Ming and Ma, Xiaojuan},
title = {Who Should I Trust: AI or Myself? Leveraging Human and AI Correctness Likelihood to Promote Appropriate Trust in AI-Assisted Decision-Making},
year = {2023},
isbn = {9781450394215},
publisher = {ACM},
address = {New York, NY, USA},
url = {https://doi.org/10.1145/3544548.3581058},
doi = {10.1145/3544548.3581058},
booktitle = {Proceedings of the 2023 CHI Conference on Human Factors in Computing Systems},
articleno = {759},
pages = {19},
location = {Hamburg, Germany},
series = {CHI '23}
}

@ARTICLE{Scharowski2023FeatureImportanceCounterfactualsExplanations,
AUTHOR={Scharowski, Nicolas and Perrig, Sebastian A. C. and Svab, Melanie and Opwis, Klaus and Brühlmann, Florian},  
TITLE={Exploring the effects of human-centered AI explanations on trust and reliance},      
JOURNAL={Frontiers in Computer Science},      	
VOLUME={5},           	
YEAR={2023},      	  
URL={https://www.frontiersin.org/articles/10.3389/fcomp.2023.1151150},       	
DOI={10.3389/fcomp.2023.1151150},      	
ISSN={2624-9898}
}

@article{Celar2023CounterfactualCausal,
author = "Celar, Lenart and Byrne, Ruth",
year = "2023",
month = "03",
pages = "",
title = "How people reason with counterfactual and causal explanations for Artificial Intelligence decisions in familiar and unfamiliar domains",
volume = "51",
journal = "Memory \& Cognition",
doi = "10.3758/s13421-023-01407-5"
}

@inproceedings{Cau2023LogicalReasoningStock,
author = {Cau, Federico Maria and Hauptmann, Hanna and Spano, Lucio Davide and Tintarev, Nava},
title = {Supporting High-Uncertainty Decisions through AI and Logic-Style Explanations},
year = {2023},
isbn = {9798400701061},
publisher = {ACM},
address = {New York, NY, USA},
url = {https://doi.org/10.1145/3581641.3584080},
doi = {10.1145/3581641.3584080},
booktitle = {Proceedings of the 28th International Conference on Intelligent User Interfaces},
pages = {251–263},
numpages = {13},
location = {Sydney, NSW, Australia},
series = {IUI '23}
}

@ARTICLE{Teso2023InteractiveExplanations,
  title    = "Leveraging explanations in interactive machine learning: An
              overview",
  author   = "Teso, Stefano and Alkan, {\"O}znur and Stammer, Wolfgang and
              Daly, Elizabeth",
    journal  = "Frontiers in Artificial Intelligence",
  volume   =  6,
  pages    = "1066049",
  month    =  feb,
  year     =  2023,
  language = "en",
doi = {https://doi.org/10.3389/frai.2023.1066049}
}

@article{Espejel2025LLMs,
title = {Low-cost language models: Survey and performance evaluation on Python code generation},
journal = {Engineering Applications of Artificial Intelligence},
volume = {140},
pages = {109490},
year = {2025},
issn = {0952-1976},
doi = {https://doi.org/10.1016/j.engappai.2024.109490},
url = {https://www.sciencedirect.com/science/article/pii/S0952197624016488},
author = {Jessica {López Espejel} and Mahaman Sanoussi {Yahaya Alassan} and Merieme Bouhandi and Walid Dahhane and El Hassane Ettifouri}
}

@Article{Zhang2024LLMsMedicalOpenCommercial,
author={Zhang, Gongbo
and Jin, Qiao
and Zhou, Yiliang
and Wang, Song
and Idnay, Betina
and Luo, Yiming
and Park, Elizabeth
and Nestor, Jordan G.
and Spotnitz, Matthew E.
and Soroush, Ali
and Campion, Thomas R.
and Lu, Zhiyong
and Weng, Chunhua
and Peng, Yifan},
title={Closing the gap between open source and commercial large language models for medical evidence summarization},
journal={NPJ Digital Medicine},
year={2024},
month={Sep},
day={09},
volume={7},
number={1},
pages={239},
issn={2398-6352},
doi={10.1038/s41746-024-01239-w},
url={https://doi.org/10.1038/s41746-024-01239-w}
}

@ARTICLE{Adams2024LlamaProprietaryRadiology,
  title     = "Llama 3 challenges proprietary state-of-the-art large language
               models in radiology board-style examination questions",
  author    = "Adams, Lisa C and Truhn, Daniel and Busch, Felix and Dorfner,
               Felix and Nawabi, Jawed and Makowski, Marcus R and Bressem, Keno
               K",
  journal   = "Radiology",
  publisher = "Radiological Society of North America (RSNA)",
  volume    =  312,
  number    =  2,
  pages     = "e241191",
  month     =  aug,
  year      =  2024,
  language  = "en"
}

@inproceedings{Shashidhar2023LLMsOpenSource,
    title = "Democratizing {LLM}s: An Exploration of Cost-Performance Trade-offs in Self-Refined Open-Source Models",
    author = "Shashidhar, Sumuk  and
      Chinta, Abhinav  and
      Sahai, Vaibhav  and
      Wang, Zhenhailong  and
      Ji, Heng",
    editor = "Bouamor, Houda  and
      Pino, Juan  and
      Bali, Kalika",
    booktitle = "Findings of the Association for Computational Linguistics: EMNLP 2023",
    month = dec,
    year = "2023",
    address = "Singapore",
    publisher = "Association for Computational Linguistics",
    url = "https://aclanthology.org/2023.findings-emnlp.608/",
    doi = "10.18653/v1/2023.findings-emnlp.608",
    pages = "9070--9084"
}

@InProceedings{Panwar2024ChatgptLlama,
author="Panwar, Shivam
and Bansal, Anukriti
and Zareen, Farhana",
editor="Sharma, Harish
and Shrivastava, Vivek
and Tripathi, Ashish Kumar
and Wang, Lipo",
title="Comparative Analysis of Large Language Models for Question Answering from Financial Documents",
booktitle="Communication and Intelligent Systems",
year="2024",
publisher="Springer Nature Singapore",
address="Singapore",
pages="297--308",
isbn="978-981-97-2079-8"
}

@article{landis1977application,
  title={An application of hierarchical kappa-type statistics in the assessment of majority agreement among multiple observers},
  author={Landis, J Richard and Koch, Gary G},
  journal={Biometrics},
  pages={363--374},
  year={1977},
  publisher={JSTOR}
}

@inproceedings{buono2023warnings,
author = {Buono, Paolo and Desolda, Giuseppe and Greco, Francesco and Piccinno, Antonio},
title = {Let warnings interrupt the interaction and explain: designing and evaluating phishing email warnings},
year = {2023},
isbn = {9781450394222},
publisher = {ACM},
address = {New York, NY, USA},
url = {https://doi.org/10.1145/3544549.3585802},
doi = {10.1145/3544549.3585802},
booktitle = {Extended Abstracts of the 2023 CHI Conference on Human Factors in Computing Systems},
articleno = {197},
numpages = {6},
pages = {1--6},
location = {Hamburg, Germany},
series = {CHI EA '23}
}

@article{chen2021generate,
  title={Generate natural language explanations for recommendation},
  author={Chen, Hanxiong and Chen, Xu and Shi, Shaoyun and Zhang, Yongfeng},
  journal={arXiv preprint arXiv:2101.03392},
  year={2021}
}

@inproceedings{Liao2020CHI,
  author = {Liao, Q. Vera and Gruen, Daniel and Miller, Stephen},
  title = {Questioning the AI: Informing Design Practices for Explainable AI User Experiences},
  booktitle = {Proceedings of the 2020 CHI Conference on Human Factors in Computing Systems},
  year = {2020},
  pages = {1--15},
  doi = {10.1145/3313831.3376590}
}

@article{chen2010big,
  title={How big is a big odds ratio? Interpreting the magnitudes of odds ratios in epidemiological studies},
  author={Chen, Henian and Cohen, Patricia and Chen, Sophie},
  journal={Communications in Statistics—simulation and Computation{\textregistered}},
  volume={39},
  number={4},
  pages={860--864},
  year={2010},
  publisher={Taylor \& Francis}
}

@incollection{wogalter2018communication,
  title={Communication-human information processing (C-HIP) model},
  author={Wogalter, Michael S},
  booktitle={Forensic human factors and ergonomics},
  pages={33--49},
  year={2018},
  publisher={CRC Press}
}

@article{amran2018habituation,
  title={Habituation effects in computer security warning},
  author={Amran, Ammar and Zaaba, Zarul Fitri and Mahinderjit Singh, Manmeet Kaur},
  journal={Information security journal: A global perspective},
  volume={27},
  number={4},
  pages={192--204},
  year={2018},
  publisher={Taylor \& Francis}
}

@book{Kahneman2011Thinking,
  author = {Kahneman, Daniel},
  title = {Thinking, Fast and Slow},
  publisher = {Farrar, Straus and Giroux},
  year = {2011}
}

@inproceedings{sheahan2024designing,
author = {Sheahan, Jacob and Chatting, David and Collins, Robert and Bley, Jessica and Eriksson, Alexander and Taylor, Nick and Rozendaal, Marco C.},
title = {Designing with Friction: Inverting Notions of Seamless Technology},
year = {2024},
isbn = {9798400709654},
publisher = {Association for Computing Machinery},
address = {New York, NY, USA},
url = {https://doi.org/10.1145/3677045.3685504},
doi = {10.1145/3677045.3685504},
booktitle = {Adjunct Proceedings of the 2024 Nordic Conference on Human-Computer Interaction},
articleno = {59},
numpages = {4},
location = {Uppsala, Sweden},
series = {NordiCHI '24 Adjunct}
}

@article{zhao2024explainability,
  author = {Zhao, Haiyan and Chen, Hanjie and Yang, Fan and Liu, Ninghao and Deng, Huiqi and Cai, Hengyi and Wang, Shuaiqiang and Yin, Dawei and Du, Mengnan},
title = {Explainability for Large Language Models: A Survey},
year = {2024},
issue_date = {April 2024},
publisher = {ACM},
address = {New York, NY, USA},
volume = {15},
number = {2},
issn = {2157-6904},
url = {https://doi.org/10.1145/3639372},
doi = {10.1145/3639372},
journal = {ACM Transactions on Intelligent Systems and Technology},
month = feb,
articleno = {20},
numpages = {38}
}

@inproceedings{gupta2014emerging,
  author={Gupta, Srishti and Kumaraguru, Ponnurangam},
  booktitle={2014 APWG Symposium on Electronic Crime Research (eCrime)}, 
  title={Emerging phishing trends and effectiveness of the anti-phishing landing page}, 
  year={2014},
  volume={},
  number={},
  pages={36-47},
  doi={10.1109/ECRIME.2014.6963163},
  url = {https://doi.org/10.1109/ECRIME.2014.6963163}
}

@article{elassal2020benchmarking,
   author = {El Aassal, Ayman and Baki, Shahryar and Das, Avisha and Verma, Rakesh M.},
   title = {An In-Depth Benchmarking and Evaluation of Phishing Detection Research for Security Needs},
   journal = {IEEE Access},
   volume = {8},
   pages = {22170-22192},
   ISSN = {2169-3536},
   DOI = {10.1109/ACCESS.2020.2969780},
   url = {https://ieeexplore.ieee.org/document/8970564},
   year = {2020},
   type = {Journal Article}
}

@misc{sotirakopoulos2011challenges,
   author = {Sotirakopoulos, Andreas and Hawkey, Kirstie and Beznosov, Konstantin},
   title = {On the Challenges in Usable Security Lab Studies: Lessons Learned from Replicating a Study on SSL Warnings},
   publisher = {ACM},
   pages = {18},
   DOI = {10.1145/2078827.2078831},
   url = {https://doi.org/10.1145/2078827.2078831},
   year = {2011},
   type = {Conference Paper}
}

@inproceedings{matsuura2021careless,
author = {Matsuura, Tenga and Hasegawa, Ayako A. and Akiyama, Mitsuaki and Mori, Tatsuya},
title = {Careless Participants Are Essential for Our Phishing Study: Understanding the Impact of Screening Methods},
year = {2021},
isbn = {9781450384230},
publisher = {ACM},
address = {New York, NY, USA},
url = {https://doi.org/10.1145/3481357.3481515},
doi = {10.1145/3481357.3481515},
booktitle = {Proceedings of the 2021 European Symposium on Usable Security},
pages = {36–47},
numpages = {12},
location = {Karlsruhe, Germany},
series = {EuroUSEC '21}
}

@article{RN3,
   author = {Vilone, Giulia and Longo, Luca},
   title = {Notions of explainability and evaluation approaches for explainable artificial intelligence},
   journal = {Information Fusion},
   volume = {76},
   number = {C},
   pages = {89-106},
   ISSN = {1566-2535},
   DOI = {10.1016/j.inffus.2021.05.009},
   url = {https://doi.org/10.1016/j.inffus.2021.05.009},
   year = {2021},
   type = {Journal Article}
}

@inproceedings{egelman2008warned,
author = {Egelman, Serge and Cranor, Lorrie Faith and Hong, Jason},
title = {You've been warned: an empirical study of the effectiveness of web browser phishing warnings},
year = {2008},
isbn = {9781605580111},
publisher = {ACM},
address = {New York, NY, USA},
url = {https://doi.org/10.1145/1357054.1357219},
doi = {10.1145/1357054.1357219},
booktitle = {Proceedings of the SIGCHI Conference on Human Factors in Computing Systems},
pages = {1065–1074},
numpages = {10},
location = {Florence, Italy},
series = {CHI '08}
}

@article{wogalter2002research,
   author = {Wogalter, Michael S. and Conzola, Vincent C. and Smith-Jackson, Tonya L.},
   title = {Research-based guidelines for warning design and evaluation},
   journal = {Applied Ergonomics},
   volume = {33},
   number = {3},
   pages = {219-230},
   ISSN = {0003-6870},
   DOI = {https://doi.org/10.1016/S0003-6870(02)00009-1},
   url = {https://www.sciencedirect.com/science/article/pii/S0003687002000091},
   year = {2002},
   type = {Journal Article}
}

@inproceedings{wu2006security,
author = {Wu, Min and Miller, Robert C. and Garfinkel, Simson L.},
title = {Do security toolbars actually prevent phishing attacks?},
year = {2006},
isbn = {1595933727},
publisher = {ACM},
address = {New York, NY, USA},
url = {https://doi.org/10.1145/1124772.1124863},
doi = {10.1145/1124772.1124863},
booktitle = {Proceedings of the SIGCHI Conference on Human Factors in Computing Systems},
pages = {601–610},
numpages = {10},
location = {Montr\'{e}al, Qu\'{e}bec, Canada},
series = {CHI '06}
}

@inproceedings{petelka2019put,
author = {Petelka, Justin and Zou, Yixin and Schaub, Florian},
title = {Put Your Warning Where Your Link Is: Improving and Evaluating Email Phishing Warnings},
year = {2019},
isbn = {9781450359702},
publisher = {ACM},
address = {New York, NY, USA},
url = {https://doi.org/10.1145/3290605.3300748},
doi = {10.1145/3290605.3300748},

booktitle = {Proceedings of the 2019 CHI Conference on Human Factors in Computing Systems},
pages = {1–15},
numpages = {15},
location = {Glasgow, Scotland Uk},
series = {CHI '19}
}

@techreport{bauer2013warning,
    author = {Bauer, Lujo and Bravo-Lillo, Cristian and Cranor, Lorrie and Fragkaki, Elli},
    title = {Warning Design Guidelines},
    institution = {Carnegie Mellon University},
    year = {2013},
    url = {https://www.researchgate.net/publication/258499093_Warning_Design_Guidelines}
}

@inproceedings{bravolillo2011improving,
author = {Bravo-Lillo, Cristian and Cranor, Lorrie Faith and Downs, Julie and Komanduri, Saranga and Sleeper, Manya},
title = {Improving computer security dialogs},
year = {2011},
isbn = {9783642237676},
publisher = {Springer-Verlag},
address = {Berlin, Heidelberg},
booktitle = {Proceedings of the 13th IFIP TC 13 International Conference on Human-Computer Interaction - Volume Part IV},
pages = {18–35},
numpages = {18},
location = {Lisbon, Portugal},
series = {INTERACT'11},
url = {https://dl.acm.org/doi/10.5555/2042283.2042286},
doi = {10.5555/2042283.2042286}
}

@InProceedings{desolda2019alerting,
author="Desolda, Giuseppe
and Di Nocera, Francesco
and Ferro, Lauren
and Lanzilotti, Rosa
and Maggi, Piero
and Marrella, Andrea",
editor="Moallem, Abbas",
title="Alerting Users About Phishing Attacks",
booktitle="HCI for Cybersecurity, Privacy and Trust",
year="2019",
publisher="Springer International Publishing",
address="Cham",
pages="134--148",
isbn="978-3-030-22351-9",
url={https://link.springer.com/chapter/10.1007/978-3-030-22351-9_9},
doi={10.1007/978-3-030-22351-9_9}
}

@article{kim2009habituation,
   author = {Kim, Soyun and Wogalter, Michael S.},
   title = {Habituation, Dishabituation, and Recovery Effects in Visual Warnings},
   journal = {Human Factors and Ergonomics Society Annual Meeting},
   volume = {53},
   number = {20},
   pages = {1612-1616},
   DOI = {10.1177/154193120905302015},
   url = {https://journals.sagepub.com/doi/abs/10.1177/154193120905302015},
   year = {2009},
   type = {Journal Article}
}

@inproceedings{akhawe2013alice,
author = {Akhawe, Devdatta and Felt, Adrienne Porter},
title = {Alice in warningland: a large-scale field study of browser security warning effectiveness},
year = {2013},
isbn = {9781931971034},
publisher = {USENIX Association},
address = {USA},
booktitle = {Proceedings of the 22nd USENIX Conference on Security},
pages = {257–272},
numpages = {16},
location = {Washington, D.C.},
series = {SEC'13},
url = {https://dl.acm.org/doi/10.5555/2534766.2534789},
doi = {10.5555/2534766.2534789}
}

@misc{anderson2015polymorphic,
   author = {Anderson, Bonnie Brinton and Kirwan, C. Brock and Jenkins, Jeffrey L. and Eargle, David and Howard, Seth and Vance, Anthony},
   title = {How Polymorphic Warnings Reduce Habituation in the Brain: Insights from an fMRI Study},
   publisher = {ACM},
   pages = {2883-–2892},
   month = {Apr 2015},
   DOI = {10.1145/2702123.2702322},
   url = {https://doi.org/10.1145/2702123.2702322},
   year = {2015},
   type = {Conference Paper}
}

@misc{saxena2018precision,
   author = {Saxena, Shruti},
   title = {Precision vs Recall},
   month = {13 Feb. 2024},
   url = {https://medium.com/@shrutisaxena0617/precision-vs-recall-386cf9f89488},
   year = {2018},
   type = {Electronic Article}
}

@article{kumaraguru2010teaching,
   author = {Kumaraguru, Ponnurangam and Sheng, Steve and Acquisti, Alessandro and Cranor, Lorrie F. and Hong, Jason},
   title = {Teaching Johnny not to fall for phish},
   journal = {ACM Transactions on Internet Technology},
   volume = {10},
   number = {2},
   pages = {1-31},
   ISSN = {1533-5399},
   DOI = {10.1145/1754393.1754396},
   url = {https://doi.org/10.1145/1754393.1754396},
   year = {2010},
   type = {Journal Article}
}

@misc{ibm2024security,
   author = {IBM},
   title = {Security X-Force Threat Intelligence Index},
   month = {14 May 2024},
   url = {https://www.ibm.com/reports/threat-intelligence},
   year = {2024},
   type = {Electronic Article}
}

@article{desolda2021human,
author = {Desolda, Giuseppe and Ferro, Lauren S. and Marrella, Andrea and Catarci, Tiziana and Costabile, Maria Francesca},
title = {Human Factors in Phishing Attacks: A Systematic Literature Review},
year = {2021},
issue_date = {November 2022},
publisher = {ACM},
address = {New York, NY, USA},
volume = {54},
number = {8},
issn = {0360-0300},
url = {https://doi.org/10.1145/3469886},
doi = {10.1145/3469886},
journal = {ACM Comput. Surv.},
month = oct,
articleno = {173},
numpages = {35}
}

@inproceedings{gholampour2023adversarial,
author = {Mehdi Gholampour, Parisa and Verma, Rakesh M.},
title = {Adversarial Robustness of Phishing Email Detection Models},
year = {2023},
isbn = {9798400700996},
publisher = {ACM},
address = {New York, NY, USA},
url = {https://doi.org/10.1145/3579987.3586567},
doi = {10.1145/3579987.3586567},
booktitle = {Proceedings of the 9th ACM International Workshop on Security and Privacy Analytics},
pages = {67–76},
numpages = {10},
location = {Charlotte, NC, USA},
series = {IWSPA '23}
}

@article{Guidotti2019Survey,
  title = {A {{Survey}} of {{Methods}} for {{Explaining Black Box Models}}},
  author = {Guidotti, Riccardo and Monreale, Anna and Ruggieri, Salvatore and Turini, Franco and Giannotti, Fosca and Pedreschi, Dino},
  year = {2019},
  month = sep,
  journal = {ACM Computing Surveys},
  volume = {51},
  number = {5},
  pages = {1--42},
  issn = {0360-0300, 1557-7341},
  doi = {10.1145/3236009},
  url = {https://dl.acm.org/doi/10.1145/3236009},
  urldate = {2024-03-12},
  langid = {english}
}

@article{Gunning2019DARPAs,
  title = {{{DARPA}}'s {{Explainable Artificial Intelligence Program}}},
  author = {Gunning, David and Aha, David W.},
  year = {2019},
  month = jun,
  journal = {AI Magazine},
  volume = {40},
  number = {2},
  pages = {44--58},
  issn = {0738-4602, 2371-9621},
  doi = {10.1609/aimag.v40i2.2850},
  url = {https://onlinelibrary.wiley.com/doi/10.1609/aimag.v40i2.2850},
  urldate = {2024-12-04},
  copyright = {http://onlinelibrary.wiley.com/termsAndConditions\#vor},
  langid = {english}
}

@misc{olmstead2017securityquestionnaire,
   author = {Olmstead, Kenneth and Smith, Aaron},
   title = {What the Public Knows About Cybersecurity},
   publisher = {Pew Research Center},
   number = {19/11/2024},
   url = {https://www.pewresearch.org/internet/2017/03/22/what-the-public-knows-about-cybersecurity/},
   year = {2017},
   type = {Web Page}
}

@misc{hart1986NASATLX,
   author = {Hart, Sandra G.},
   title = {NASA Task Load Index (NASA-TLX)},
   number = {30/05/2024},
   url = {https://ntrs.nasa.gov/api/citations/20000021488/downloads/20000021488.pdf},
   year = {1986},
   type = {Web Page}
}

@article{distler2021slr,
   author = {Distler, Verena and Fassl, Matthias and Habib, Hana and Krombholz, Katharina and Lenzini, Gabriele and Lallemand, Carine and Cranor, Lorrie Faith and Koenig, Vincent},
   title = {A Systematic Literature Review of Empirical Methods and Risk Representation in Usable Privacy and Security Research},
   journal = {ACM Transactions on Computer-Human Interaction},
   volume = {28},
   number = {6},
   pages = {1-50},
   ISSN = {1073-0516},
   DOI = {10.1145/3469845},
   url = {https://doi.org/10.1145/3469845},
   year = {2021},
   type = {Journal Article}
}

@article{coelho2020needforcognition6item,
author = {Lins de Holanda Coelho, Gabriel and Hanel, Paul H. P. and Wolf, Lukas J.},
title ={The Very Efficient Assessment of Need for Cognition: Developing a Six-Item Version},
journal = {Assessment},
short = {ASM},
volume = {27},
number = {8},
pages = {1870-1885},
year = {2020},
doi = {10.1177/1073191118793208},
note ={PMID: 30095000},
URL = {https://doi.org/10.1177/1073191118793208},
eprint = { https://doi.org/10.1177/1073191118793208}
}

@article{cacioppo1982needforcognition,
  title={The need for cognition},
  author={Cacioppo, John T and Petty, Richard E},
  journal={Journal of personality and social psychology},
  volume={42},
  number={1},
  pages={116},
  year={1982},
  publisher={American Psychological Association},
  url = {https://psycnet.apa.org/doi/10.1037/0022-3514.42.1.116},
  doi = {10.1037/0022-3514.42.1.116}
}

@article{braun2006thematic,
   author = {Braun, Virginia and Clarke, Victoria},
   title = {Using thematic analysis in psychology},
   journal = {Qualitative Research in Psychology},
   volume = {3},
   number = {2},
   pages = {77-101},
   ISSN = {1478-0895(Electronic),1478-0887(Print)},
   DOI = {10.1191/1478088706qp063oa},
   url = {https://www.tandfonline.com/doi/abs/10.1191/1478088706qp063oa},
   year = {2006},
   type = {Journal Article}
}

@article{greitzer2021experimental,
  title={Experimental investigation of technical and human factors related to phishing susceptibility},
  author={Greitzer, Frank L and Li, Wanru and Laskey, Kathryn B and Lee, James and Purl, Justin},
  journal={ACM Transactions on Social Computing},
  volume={4},
  number={2},
  pages={1--48},
  year={2021},
  publisher={ACM New York, NY, USA},
url = {https://doi.org/10.1145/3461672},
doi = {10.1145/3461672},
}

@article{Cacioppo1996NFCCognitive,
author = {Cacioppo, John and Petty, Richard and Feinstein, Jeffrey and Jarvis, Blair},
year = {1996},
month = {03},
pages = {197-253},
title = {Dispositional Differences in Cognitive Motivation: The Life and Times of Individuals Varying in Need for Cognition},
volume = {119},
journal = {Psychological Bulletin},
doi = {10.1037/0033-2909.119.2.197}
}

@article{Bucinca2021NFC,
author = {Bu\c{c}inca, Zana and Malaya, Maja Barbara and Gajos, Krzysztof Z.},
title = {To Trust or to Think: Cognitive Forcing Functions Can Reduce Overreliance on AI in AI-assisted Decision-making},
year = {2021},
issue_date = {April 2021},
publisher = {ACM},
address = {New York, NY, USA},
volume = {5},
number = {CSCW1},
url = {https://doi.org/10.1145/3449287},
doi = {10.1145/3449287},
journal={Proceedings of the ACM on Human-computer Interaction},
pages={1--21}
}

@inproceedings{Gajos2022NFC,
author = {Gajos, Krzysztof Z. and Mamykina, Lena},
title = {Do People Engage Cognitively with AI? Impact of AI Assistance on Incidental Learning},
year = {2022},
isbn = {9781450391443},
publisher = {ACM},
address = {New York, NY, USA},
url = {https://doi.org/10.1145/3490099.3511138},
doi = {10.1145/3490099.3511138},
booktitle = {Proceedings of the 27th International Conference on Intelligent User Interfaces},
pages = {794–806},
numpages = {13},
location = { Helsinki, Finland},
series = {IUI '22}
}

@misc{Bucinca2024Optimizing,
      title={Towards Optimizing Human-Centric Objectives in AI-Assisted Decision-Making With Offline Reinforcement Learning}, 
      author={Zana Bu\c{c}inca and Siddharth Swaroop and Amanda E. Paluch and Susan A. Murphy and Krzysztof Z. Gajos},
      year={2024},
      eprint={2403.05911},
      archivePrefix={arXiv},
      primaryClass={cs.HC},
      url={https://arxiv.org/abs/2403.05911}, 
}

@inproceedings{ViganoMagazzeni2020,
  author       = {Luca Vigan{\`{o}} and
                  Daniele Magazzeni},
  title        = {Explainable Security},
  booktitle    = {{IEEE} European Symposium on Security and Privacy Workshops, EuroS{\&}P
                  Workshops 2020, Genoa, Italy, September 7-11, 2020},
  pages        = {293--300},
  publisher    = {{IEEE}},
  year         = {2020},
}

@article{Kirk1996,
author = {Roger E. Kirk},
title ={Practical Significance: A Concept Whose Time Has Come},

journal = {Educational and Psychological Measurement},
volume = {56},
number = {5},
pages = {746-759},
year = {1996},
doi = {10.1177/0013164496056005002},

URL = {https://doi.org/10.1177/0013164496056005002}
}

@article{fritz2012effect,
  title={Effect size estimates: current use, calculations, and interpretation},
  author={Fritz, Catherine O and Morris, Peter E and Richler, Jennifer J},
  journal={Journal of experimental psychology: General},
  volume={141},
  number={1},
  pages={2},
  year={2012},
  publisher={American Psychological Association},
  doi = {10.1037/a0024338}
}

@article{djatsa2020threat,
  title={Threat Perceptions, Avoidance Motivation and Security Behaviors Correlations},
  author={Djatsa, Fabrice},
  journal={Journal of Information Security},
  volume={11},
  number={01},
  pages={19},
  year={2020},
  publisher={Scientific Research Publishing},
  doi = {10.4236/jis.2020.111002}
}

@Article{Cau2025HCXAILoan,
author={Cau, Federico Maria
and Spano, Lucio Davide},
title={Exploring the impact of explainable AI and cognitive capabilities on users' decisions},
journal={User Modeling and User-Adapted Interaction},
year={2025},
month={Dec},
day={06},
volume={36},
number={1},
pages={3},
issn={1573-1391},
doi={10.1007/s11257-025-09438-0},
url={https://doi.org/10.1007/s11257-025-09438-0}
}

\onecolumn

\begin{appendices}

\section{A. Prompts for generating explanations}
\label{appendix:prompts}

To generate the explanations used in the study, we used a modified version of APOLLO~\cite{desolda2025apollo}, an LLM-based tool for phishing email classification and explanation. While the full details of APOLLO are reported in the original paper~\cite{desolda2025apollo}, for the sake of auto-consistency, we provide the prompts that the system uses in the following.

\subsection{Email classification prompt}

\begin{lstlisting}[caption={First prompt. This makes the LLM generate a classification outcome for an email and generate a first, raw explanation.}]
    You are a cybersecurity and human-computer interaction expert who has the goal of detecting if an email is legitimate or phishing and helping the user understand why a specific email is dangerous (or genuine) in order to make more informed decisions.
    The user will submit the email (headers + subject + body) optionally accompanied by information on the URLs in the email as:
    - server location;
    - VirusTotal scans reporting the number of scanners that detected the URL as harmless, undetected, malicious.
    Your goal is to output a JSON object containing:
    - The classification result (label).
    - The probability as a percentage of the email being phishing (0%=email is surely legitimate, 100%=email is surely phishing) (phishing_probability).
    - A list of persuasion principles that were applied by the alleged attacker (if any); each persuasion principle should be an object containing:
        the persuasion principle name (authority, scarcity, etc.),
        the part of the email that makes you say that the persuasion principle is being applied;
        a brief rationale for each principle.
    - A list of 3 to 5 features that could indicate the danger (or legitimacy) of the email; the explanations must be understandable by users with no cybersecurity or computer expertise.
    Desired format:
    label: <phishing/legit>
    phishing_probability: <0-100%>
    persuasion_principles: [array of persuasion principles, each having: (name, specific sentences, rationale)] 
    explanation: [array of 3-5 features explained]
    
    Email:
    [HEADERS] {email headers} [\HEADERS]
    [SUBJECT] {email subject} [\SUBJECT]
    [BODY]
    {email body}
    [\BODY]
    URL Information:
    Server location: {URL geolocation}
    VirusTotal scan: [
    harmless: {n_harmless}, 
    undetected: {n_undetected}, 
    malicious: {n_malicious} 
    ]
\end{lstlisting}

\subsection{Explanation generation prompts}

The results of the email classification are then sent to the LLM, along with a second prompt, adapted from APOLLO~\cite{desolda2025apollo}, which varies according to the explanation type (feature-based or counterfactual). 

\begin{lstlisting}[caption={This prompt generates a refined feature-based explanation message following a specific structure. Note that, for the aims of this study, the model is primed to explain a specific feature injected in the prompt ("feature to explain").}]
Now consider the [feature to explain] feature and construct a brief explanation message construct a brief explanation message (max 50 words) directed to naive users (with no knowledge of cybersecurity) that will follow this structure:
1. description of the most relevant phishing feature
2. explanation of the hazard
3. consequences of a successful phishing attack
For example, an explanation that explains that the domain of a website is suspiciously young would be:
"The URL in the email leads to a website created N days ago. Young websites are famous for criminal activity. There is a potential risk if you proceed."
Another example of explaining that the email is suspicious because of too many special characters in its body would be:
"Many special characters have been detected in the email. Malicious people use them to disguise text and deceive you. Your data could be stolen."

Desired format:
[description of the feature]. [hazard explanation]. [consequences of a successful attack].

\end{lstlisting}

\begin{lstlisting}[caption={This prompt generates a refined counterfactual explanation message following a specific structure. Note that, for the aims of this study, the model is primed to explain a specific feature injected in the prompt ("feature to explain").}]
Now consider the [feature to explain] feature and construct a brief counterfactual explanation message (max 50 words) directed to naive users (with no knowledge of cybersecurity or computers) that will follow this structure: 
      1. description of the phishing feature
      2. hypothetical conditions for which it would have been safe
      3. counter-example of a safe feature
Avoid using technical jargon such as "IP address", "domain", "URL", etc. It might also be useful to display the URL within the warning message.

For example, an explanation that explains that the domain of a website is suspiciously young would be:
"The URL in the email leads to a website created N days ago. The email would have been considered safe if the website were older than a typical attack cycle, as this timeframe provides a verifiable reputation. A safe link might point to a domain with years of established activity."

Another example of explaining that the email is suspicious because of too many special characters in its body would be:
"Many special characters have been detected in the email. The email would have been considered safe if it mostly contained standard text (letters and numbers), as this eliminates the possibility of visually deceptive text used to trick users. A safe email would only use characters typically found in formal correspondence."

Desired format:
[description of the feature]. [hypothetical safety conditions]. [counter-example]

\end{lstlisting}

\section{B. Warning Messages}
\label{appendix:warnings}

Below are the explanation messages contained in the warnings for each condition. Each condition had three warnings, one per email: \textit{IP address}, \textit{Top-level domain mispositioned}, and \textit{Link mismatch}. 

\subsection{Claude feature-based explanations}

\begin{itemize}
    \item \textit{IP address}. The link shows numbers (92.233.24.33) instead of ``instagram.com'' like real Instagram emails would use. This is a deceptive trick to send you to a fake website. If you enter your account details there, attackers will gain control of your Instagram account
    \item \textit{Top-level domain mispositioned}. The website amazonservices.com.cz is trying to look like a real Amazon page by using 'amazon' in its address. This is a trick to make you think you're visiting Amazon when you're not. If you enter your login details, scammers could take control of your real Amazon account.
    \item \textit{Link mismatch}. The link shown as 'protect your account' may take you somewhere different from what you see. Attackers use this trick to make you think you're going to Facebook when you're not. This could lead to your Facebook password being stolen.
\end{itemize}

\subsection{Llama feature-based explanations}

\begin{itemize}
    \item \textit{IP address}. The link http://92.233.24.33 is a string of numbers instead of a normal website name. This site might be fake and try to trick you. You might be giving away your private information.
    \item \textit{Top-level domain mispositioned}. The link https://amazonservices.com.cz/ account.php has a strange company name extension. This site might be pretending to be something it's not. You might be disclosing private information.
    \item \textit{Link mismatch}. The link \url{https://www.facebook.com/hacked/disavow? u=100000125023309\&nArdInDS2\&lit\_IT\&ext1548538159} is an imitation of the original Facebook link. This site might be intended to take you to a different place. You might be disclosing private information.
\end{itemize}

\subsection{Claude counterfactual explanations}

\begin{itemize}
    \item \textit{IP address}. The link in the email points to a string of numbers (92.233. 24.33/instagram/login.php) instead of the official Instagram website name. The email would have been safe if the link used Instagram's actual website name, which helps users verify they're going to the real Instagram site. A safe link would have looked like: https://instagram.com /account/reset
    \item \textit{Top-level domain mispositioned}. The link 'amazonservices.com.cz' tries to trick you by putting '.com' in the middle instead of at the end. The email would have been considered safe if the website address ended with '.com' or matched Amazon's official website format. A safe link from Amazon would look like: https://amazon.com/account or https://amaz on.it/account
    \item \textit{Link mismatch}. The email shows a safe link but leads to 'facebook.com/hacked'. The email would have been safe if the link you see matched exactly where it takes you, which helps ensure you're going to the real Facebook website. A safe link would show and lead to the same place, like: facebook.com/help/security
\end{itemize}

\subsection{Llama counterfactual explanations}

\begin{itemize}
    \item \textit{IP address}. The link http://92.233.24.33 is a string of numbers instead of a normal website name. The email would have been considered safe if it used a recognisable website name, like a company's official name, which helps verify the link's authenticity. A safe link might have looked like: https://instagram.com/login
    \item \textit{Top-level domain mispositioned}. The link \url{https://amazonservices.com. cz/account.php} has a suspicious web address with a ``.com'' in the wrong place. The email could have been considered safe if the ``.com'' was at the end of the address. A safe link might have looked like: https://amazon.com/account-update
    \item \textit{Link mismatch}. The link text ``protect your account'' hides the actual link \url{https://www.facebook.com/hacked/disavow?u=100000125023309 \&nArdInDS2\&lit\_IT\&ext1548538159}. The email would have been considered safe if the link text matched the actual link, which helps ensure the link is trustworthy and not trying to trick you. A safe link might have looked like: https://www.facebook.com/account-security.
\end{itemize}

\subsection{Baseline - Manually-generated feature-based explanations}

\begin{itemize}
    \item \textit{IP address}. Usually, websites use the URL instead of the IP address to make it easier for you to browse the web. However, an IP address was found in the email. Similar emails are harmful and steal private information. There is a potential risk of being cheated if you proceed.”
    \item \textit{Top-level domain mispositioned}. In the URL present in the email (\url{https://amazonservices.com.cz/ account.php}), the top-level domain (e.g., “.com“) is in an abnormal position. This could indicate that the URL leads to a fake website. Such websites might steal your personal information.
    \item \textit{Link mismatch}. This email reports a link that is different from the actual one \url{https://www.facebook.com/hacked/disavow? u=1000001250 23309\&nArdInDS2\&lit\_IT\&ext1548538159}. This site might be intended to take you to a different place. You might be disclosing private information.
\end{itemize}

\section{C. Logistic Regression Detailed Results}
\label{appendix:lr}

\begin{longtable}{@{}p{2.8cm}llccc@{}}
\caption{Logistic regression results on CTR (click-through rate). OR$>$1 increases click likelihood; OR$<$1 decreases it.}
\label{tab:user_factors_ctr} \\
\toprule
\textbf{Variable} & \textbf{Cond.} & \textbf{Emails} & \textbf{OR} & \textbf{$p$} & \textbf{Size} \\
\midrule
\endfirsthead
\multicolumn{6}{@{}l}{\small (continued)} \\
\toprule
\textbf{Variable} & \textbf{Cond.} & \textbf{Emails} & \textbf{OR} & \textbf{$p$} & \textbf{Size} \\
\midrule
\endhead
\midrule
\multicolumn{6}{r}{\small (continued on next page)} \\
\endfoot
\bottomrule
\endlastfoot

\textbf{Familiarity} & CF agg & ALL & 2.20 & .001 & Large \\
 & CF agg & TP & 2.34 & .001 & Large \\
 & CF agg & FP & 2.53 & .012 & Large \\
 & Claude agg & TP & 1.83 & .028 & Large \\
 & Claude CF & ALL & 2.23 & .033 & Large \\
 & Claude CF & TP & 2.69 & .022 & Large \\
 & Claude CF & FP & 3.24 & .050 & Large \\
 & Llama CF & ALL & 2.14 & .021 & Large \\
 & Llama CF & TP & 2.05 & .030 & Large \\
\midrule

\textbf{Don't continue} & CF agg & ALL & 0.21 & $<$.001 & Large \\
 & CF agg & TP & 0.18 & $<$.001 & Large \\
 & NoLLM & ALL & 0.14 & $<$.001 & Large \\
 & NoLLM & TP & 0.10 & $<$.001 & Large \\
 & Llama agg & ALL & 0.26 & .002 & Large \\
 & Llama agg & TP & 0.23 & .001 & Large \\
 & Claude agg & ALL & 0.41 & .034 & Large \\
 & Claude agg & TP & 0.34 & .023 & Large \\
\midrule

\textbf{Female} & CF agg & ALL & 0.006 & .023 & Large \\
 & CF agg & FP & 0.005 & .003 & Large \\
 & CF agg & TP & 0.041 & .012 & Large \\
 & NoLLM & FP & 0.0003 & .001 & Large \\
 & NoLLM & TP & 0.0038 & $<$.001 & Large \\
 & Llama agg & FP & 0.03 & .011 & Large \\
 & Claude agg & TP & 0.035 & .022 & Large \\
 & Claude CF & ALL & 0.012 & .023 & Large \\
 & Llama CF & FP & 0.010 & .041 & Large \\
\midrule

\textbf{Male} & CF agg & TP & 2.62 & .020 & Large \\
 & Llama agg & TP & 2.13 & .037 & Large \\
 & NoLLM & FP & 0.13 & .036 & Large \\
\midrule

\textbf{NASA-TLX} & FB agg & FP & 0.52 & .004 & Moderate \\
 & Llama agg & ALL & 0.63 & .008 & Moderate \\
 & Llama agg & TP & 0.64 & .013 & Moderate \\
 & CF agg & ALL & 0.65 & .012 & Moderate \\
 & CF agg & TP & 0.64 & .012 & Moderate \\
 & Claude agg & ALL & 0.74 & .050 & Moderate \\
 & Claude agg & FP & 0.61 & .018 & Moderate \\
 & Claude FB & FP & 0.47 & .033 & Large \\
 & Llama FB & FP & 0.45 & .036 & Large \\
\midrule

\textbf{Understandability} & CF agg & ALL & 0.53 & .044 & Moderate \\
 & CF agg & TP & 0.51 & .042 & Moderate \\
 & Claude agg & FP & 0.48 & .046 & Large \\
\midrule

\textbf{Trust warning} & Llama FB & FP & 0.44 & .038 & Large \\
\midrule

\textbf{Avg hours/day} & NoLLM & ALL & 1.24 & .024 & Small \\
 & NoLLM & TP & 1.22 & .006 & Small \\
\midrule

\textbf{Age × Hours} & FB agg & ALL & 1.007 & .025 & Negl. \\
 & CF agg & ALL & 1.007 & .025 & Negl. \\
 & Llama agg & ALL & 1.007 & .025 & Negl. \\
 & Claude agg & ALL & 1.007 & .025 & Negl. \\
 & NoLLM & TP & 1.02 & .027 & Negl. \\
 & Claude CF & FP & 0.93 & .014 & Negl. \\
\midrule

\textbf{Expertise × Hours} & Claude FB & ALL & 1.07 & .042 & Negl. \\
 & Llama CF & ALL & 1.11 & .039 & Negl. \\
\midrule

\textbf{Male × NFC} & CF agg & FP & 1.44 & .001 & Moderate \\
 & CF agg & ALL & 1.21 & .008 & Small \\
 & Claude CF & ALL & 1.33 & .023 & Small \\
 & Claude CF & FP & 2.15 & .012 & Large \\
 & Claude CF & TP & 1.32 & .024 & Small \\
 & Llama agg & ALL & 1.04 & .032 & Negl. \\
\midrule

\textbf{Expertise × NFC} & Llama agg & ALL & 1.03 & .032 & Negl. \\
\midrule

\textbf{Age × Expertise} & CF agg & ALL & 0.98 & .048 & Negl. \\
 & CF agg & FP & 0.97 & .045 & Negl. \\
\midrule

\textbf{Male × Age} & Llama agg & TP & 0.88 & .02 & Negl. \\

\end{longtable}

\begin{tablenotes}
\small
\item Effect sizes: Small (OR 1.2–1.5/0.67–0.83); Moderate (1.51–2/0.5–0.66); Large ($>$2/$<$0.5). ORs between 0.84–1.19 are Negligible (Negl.).
\end{tablenotes}

\end{appendices}

\end{document}